\documentclass[11pt, a4paper]{amsart}

\usepackage{amsmath, amssymb, amsthm}
\usepackage{graphicx}
\usepackage{subfigure}
\usepackage{color}
\usepackage[hyperindex,breaklinks]{hyperref}
\usepackage{xspace}
\usepackage{framed}
\usepackage[ruled,noline]{algorithm2e}
\usepackage[a4paper]{geometry} \usepackage{amsmath, amssymb, amsthm}
\usepackage{graphicx}
\usepackage{subfigure}
\usepackage{color}
\usepackage[square,sort,comma,numbers]{natbib}
\usepackage[hyperindex,breaklinks]{hyperref}
\usepackage{xspace}
\usepackage{framed}
\usepackage{mathtools}
\usepackage{setspace}
\newtheorem{theorem}{Theorem}[section]

\newtheorem{assumption}{Assumption}

\newtheorem{corollary}{Corollary}[section]

\newtheorem{example}{Example}
\newtheorem{lemma}{Lemma}[section]
 \newtheorem{proposition}{Proposition}[section]
\newtheorem{remark}{Remark}
\hypersetup{colorlinks = true, urlcolor = blue, linkcolor = blue,
  citecolor = red }

\newcommand{\xmath}[1]{\ensuremath{#1}\xspace}

\newcommand{\bbeta}{\xmath{\boldsymbol{\theta}}}

\newcommand{\xx}{\boldsymbol{x}}
\newcommand{\yy}{\boldsymbol{y}}

\newcommand{\ddelta}{\xmath{\boldsymbol{\delta}}}

\newcommand{\Real}{\mathbb{R}}
\newcommand{\Prob}{{P}}
\newcommand{\Expc}{\mathbb{E}}
\newcommand{\VarT}{\widetilde{\mathrm{Var}}}
\newcommand{\cis}{\hat{C}_{\beta,n}^{\textnormal{IS}}}
\newcommand{\vis}{\hat{v}_{\beta,n}^{\textnormal{IS}}}
\newcommand{\Fis}{\hat{F}_{\bbeta,n}^{\textnormal{IS}}}
\newcommand{\vsa}{\hat{v}_{\beta_0,n}}
\newcommand{\csa}{\hat{C}_{\beta_0,n}}
\usepackage{graphicx}
\usepackage{subfigure}
\usepackage{color}
\usepackage[hyperindex,breaklinks]{hyperref}
\usepackage{xspace}
\usepackage{framed}
\title[Optimizing tail risks for heavy-tailed objectives]{Optimizing tail risks using an importance sampling based extrapolation for heavy-tailed objectives}
\author[Deo]{Anand Deo}\address{Tata Institute of Fundamental Research}
\author[Murthy]{Karthyek Murthy} \address{Singapore University of Technology and Design}
\begin{document}
\maketitle
\begin{abstract}
  Motivated by the prominence of Conditional Value-at-Risk (CVaR) as a
  measure for tail risk in settings affected by uncertainty, we
  develop a new formula for approximating CVaR based optimization
  objectives and their gradients from limited samples. 
  Unlike the state-of-the-art sample average approximations which
  require impractically large amounts of data in tail probability
  regions, the proposed approximation scheme exploits the
  self-similarity of heavy-tailed distributions to extrapolate data
  from suitable lower quantiles.  The resulting approximations are
  shown to be statistically consistent and are amenable for
  optimization by means of conventional gradient descent. The
  approximation is guided by means of a systematic importance-sampling
  scheme whose asymptotic variance reduction properties are rigorously
  examined. Numerical experiments demonstrate the superiority of the
  proposed approximations and the ease of implementation points to the
  versatility of settings to which the approximation scheme can be
  applied.
\end{abstract}

\section{Introduction}
Conditional Value at Risk (CVaR) is a tail-risk measure which has
found widespread use in decision making in reliability/safety-critical
uncertain environments such as those arising in finance, operations
research, power flow distribution, motion planning, etc. (see, for
example, \cite{Ban, Krokhmal,Rockafellar, Shapiro, Tamar}).
For a random objective $L(\bbeta)$ denoting the the risk (or loss)
associated with a controllable parameter choice $\bbeta,$ the CVaR at
level $1-\beta$ is the risk averaged over the $\beta$ fraction of the
outcomes with the highest risk, as in,
\begin{align}
  C_\beta(\bbeta) := E\left[ L(\bbeta) \,\vert \, L(\bbeta) \geq v_{\beta}(\bbeta)\right],
  \label{eqdefn:Cbeta}
\end{align}
where $v_{\beta}(\bbeta)$ is the $(1-\beta)$-quantile of $L(\bbeta).$
As is often in practice, suppose that the loss structure $L(\bbeta)$
is given by $L(\bbeta) = \ell(\bbeta,\boldsymbol{X}),$ for a
deterministic function $\ell(\cdot)$ which is convex in $\bbeta$ and a
random vector $\boldsymbol{X}$ modeling uncertainty. Then the CVaR,
denoted by $C_\beta(\bbeta)$ above, is a convex risk measure with
desirable coherence properties \cite{Rockafellar}; indeed, the
convexity renders CVaR as a suitable vehicle for introducing
risk-aversion in optimization formulations.

The widespread use of CVaR in the above contexts has recently sparked
interest in its applicability towards tackling broader challenges
pertaining to fairness and reliability/safety in modern machine
learning applications: see, for e.g, \cite{Williamson,Chow}. However, despite the conceptual advantages, a key difficulty which restricts the use of CVaR in the modern data-driven applications is
the inherent statistical difficulty associated with tail risks: With
limited fraction of historical data representing extreme risk
outcomes, sample-averages possess high variance unless large amounts
of data are utilized. For example, if we wish to approximate CVaR of a
random variable within a fixed relative error at level
$1-\beta = 0.99,$ we would need at least $1/\beta = 100$ times more
samples than we would require to estimate its mean.  The impact of
insufficient data on the noisiness of mean - CVaR frontiers in the
context of portfolio optimization has been chronicled in
\cite{Caccioli18,Lim}. Due to the statistical nature of rare events,
this difficulty is present in all data-driven settings where $1-\beta$
is taken close to 1 in order to incorporate high reliability or safety
requirements.

To overcome this limitation, we consider the specific setting where
the random vector $\boldsymbol{X}$ is suitably heavy-tailed and exploit
the self-similar structure of heavy-tailed distributions to arrive at
statistically consistent approximations of CVaR-based objectives and
their gradients which can be constructed with limited samples.  Our
interest in the heavy-tailed case is motivated by the fact that tail
risks are more pronounced and risk-aversion is of paramount importance
particularly in the presence of heavy-tailed stochastic factors. Our
contributions are two-fold:
\begin{itemize}
\item[1)] Assuming oracle access to the probability density of
  $\boldsymbol{X},$ we first develop an entirely novel Importance
  Sampling (IS) scheme for approximating the CVaR objective
  $C_\beta(\bbeta)$. The IS estimator has zero bias and is shown to
  possess substantially lower variance (when compared to the naive
  sample average) in the considered setting where the use of IS, until
  now, has been well-developed only for instances where
  $\boldsymbol{X}$ is one-dimensional (or) its components are
  independent.
\item[2)] More importantly, when the probability density of
  $\boldsymbol{X}$ is not known, we show that the likelihood
  representation in the importance sampling estimator can be suitably
  approximated to result in data-driven estimators for the CVaR term
  $C_\beta(\bbeta)$ and its gradient $\nabla C_\beta(\bbeta).$
  Interestingly, the variance of the resulting approximations are only
  a fraction of those constructed from the usual sample averages and
  their bias vanish as the estimation task is made more challenging by
  letting $\beta \searrow 0.$ 
\end{itemize}
A key ingredient in verifying the statistical consistency is Theorem
\ref{thm:convergence} which establishes that the approximation,
\begin{align*}
  \nabla C_\beta(\bbeta) \approx
  \left( {\beta_0}/{\beta} \right)^{{\xi}}\nabla C_{\beta_0}(\bbeta), 
\end{align*}
for an estimable parameter ${\xi},$ has vanishing relative error for
suitably small $\beta_0,\beta$ even if $\beta_0 \gg \beta$. This is
due to the self-similarity of heavy-tailed distributions and allows
approximating CVaR at higher quantile $1 - \beta$ by means of CVaR
observed in data at lower quantile $1-\beta_0$, where $\beta_0$ is
chosen suitably larger than $\beta.$ This self-similarity phenomenon,
though well-known in extreme value theory in statistics and
quantitative risk management, has not been utilized in optimization
contexts. Our goal is to suitably facilitate the use of extreme value
theory based extrapolation in data-driven optimization by tackling the
estimation of gradient (sensitivity) of objectives with tail risks.
Moreover, among methods which specifically consider optimization in
the presence of heavy-tails, see for eg. \cite{Mainik}, the proposed
approach based on self-similarity has the added advantage that it does
not require explicit estimation of the joint dependence of
$\boldsymbol{X},$ which is a statistically challenging exercise even
in small dimensions. Numerical experiments in the context of portfolio
optimization reveal superior performance compared to using naive
sample averages even in dimensions as large as 100 (thus demonstrating
scalability) and offers reliable performance even if there is a severe
paucity of samples in the desired tail region
$L(\bbeta) \geq v_\beta(\bbeta).$

The paper is organized as follows: After introducing the importance
sampling scheme for CVaR estimation in Section \ref{sec:IS}, we use it
to derive the extrapolation based estimator in Section
\ref{sec:DD}. We demonstrate the strength of the proposed data-driven
scheme in Section \ref{sec:num-exp} by reporting results of
experiments with simulated and real datasets. We provide a sketch of
the proof of a main result in Section \ref{sec:proofs}. The entire
proofs of all the new results in the paper are presented in the appendix.


\textbf{Notation.} We use bold symbols to denote vectors, for e.g.,
$\boldsymbol{x} = (x_1,\ldots,x_d)$. For any two vectors
$\boldsymbol{a},\boldsymbol{b}$, we use $\boldsymbol{a}\boldsymbol{b}$
and $\boldsymbol{a}/\boldsymbol{b},$ respectively, to denote
component-wise multiplication and division.  
For any $a \in \Real,$ $a^+ = \max\{0,a\}$ denotes the positive part
of $a.$ The symbols $\xrightarrow{\mathcal{P}}$ and
$\xrightarrow{\mathcal{L}}$ denote convergence in probability and
distribution, respectively. For a sequence of random variables
$\{X_n\}_{n \geq 1}$, we say that $X_n$ is $o_{_\mathcal{P}}(1)$ if
$X_n\xrightarrow{\mathcal{P}} 0$.  We say that a sequence
$a_n=o(b_n)$, if as $n\to\infty$, $a_n/b_n\to 0$, and that
$a_n= \tilde{O}(b_n)$ if there exist a constant $k$ such that
$a_n/(b_n\log^k b_n)$ is asymptotically bounded.  A function
$L(\cdot)$ is slowly varying if for any constant $c,$
$L(ct)/L(t) \to 1$ as $t\to\infty.$ 
We use $\mathcal{N}(0,\mathbb{I}_d)$ to denote the $d$-dimensional
standard Gaussian vector.
%

\section{CVaR estimation via importance sampling}
\label{sec:IS}
Suppose that $L(\bbeta) := \ell(\bbeta^\intercal \boldsymbol{X})$
represents the random loss (or risk) associated with the parameter
$\bbeta \in \Theta,$ where $\Theta$ is a bounded subset of
$\mathbb{R}^d_{++},$ and $\boldsymbol{X}$ is an
$\mathbb{R}^d_+-$valued random vector whose distribution satisfies
Assumption \ref{assumption:GEV} below. The loss
$\ell: \mathbb{R} \rightarrow \mathbb{R}_+$ is such that the derivate
$\ell^\prime(u)$ exists for all $u$ exceeding a positive constant
$u_0$ and is taken to grow at most polynomially.
\begin{assumption}\label{assume:l-loss}
Suppose that $\ell^{\prime}(u) = c_1u^{\rho}(1+o(1))$, as $u\to\infty,$
for some $\rho \geq 0, c_1 > 0.$
\end{assumption}
 Commonly used losses in portfolio
optimization, newsvendor models and quadratic optimization models,
square loss, logistic loss, etc. satisfy the above assumptions.To state the assumption on the distribution of $\boldsymbol{X},$
let $\boldsymbol{M}_n$ denote the component-wise maxima of $n$
independent and identically distributed (i.i.d.)  copies of
$\boldsymbol{X}.$
\begin{assumption}
  \label{assumption:GEV}
  There exists a normalizing sequence
  $\{\boldsymbol{a}_n\}_{n \geq 1} \subseteq \mathbb{R}_d^+$ such
  that,
  as $n \rightarrow \infty,$ the sequence
  $\Vert\boldsymbol{a}_n\Vert_\infty^{-1}\boldsymbol{a}_n$ is
  convergent and the distribution of
  $\boldsymbol{M}_n/\boldsymbol{a_n}$ converges to a nondegenerate
  probability distribution.
\end{assumption}

Just like how the central limit theorem quantifies the limiting
behaviour of the i.i.d.  sum
$n^{-1/2}(\boldsymbol{X}_1 + \ldots + \boldsymbol{X}_n),$ the extreme value
theorem (see, eg. \cite{Resnickbook}) specifies all possible limiting
distributions of suitably scaled and centred maxima
$\left(\boldsymbol{M}_n - \boldsymbol{b}_n\right)/\boldsymbol{a}_n.$ 
For heavy-tailed random vectors, the centering sequence
$\{\boldsymbol{b}_n\}$ can be taken to be zero without loss of
generality, whereas $\{\boldsymbol{b}_n\}$ is necessarily divergent for
light-tailed distrubtions. Thus, Assumption \ref{assumption:GEV} is
indeed one of the general descriptions of multivariate heavy-tailed
distributions, and it includes the well-studied multivariate regularly
varying models \cite{Resnickbook}. In Assumption \ref{assumption:GEV},
the scaling constants $a_{n,1},\ldots,a_{n,d}$ capture the relative
tail heaviness of each component and the limiting distribution
describes diverse dependence distributions.

For any $\bbeta \in \Theta,$ let
$F_{\bbeta}(x) := P(\ell(\bbeta^\intercal\boldsymbol{X}) \leq x)$
denote the cumulative distribution function (c.d.f.) of the loss
$L(\bbeta).$ Given a confidence level $1-\beta \in (0,1),$
\[v_\beta(\bbeta) := F_{\bbeta}^{-1}(1-\beta) = \inf\{u \in
  \mathbb{R}: F_{\bbeta}(u) \geq 1-\beta\}\] denotes the
$(1-\beta)$-quantile of the loss
$\ell(\bbeta^\intercal\boldsymbol{X}).$ The quantile $v_\beta(\bbeta)$ is
also referred to as the Value at Risk (VaR) of the loss
$L(\bbeta) := \ell(\bbeta^\intercal\boldsymbol{X})$ at level $1-\beta.$
Then the conditional value at risk (CVaR) at level $1-\beta,$ given by
(\ref{eqdefn:Cbeta}), 
is simply the expected loss observed over the $\beta$ fraction of the
outcomes with highest loss. While both VaR and CVaR are commonly used
in practice, the use of CVaR has gained more prominence because,
unlike VaR, it quantifies the extent of extreme risk and encourages
diversification \cite{mcneil}.

\subsection{Sample-average approximation of CVaR}
\label{sec:SAA}
Suppose that $\boldsymbol{X}_1, \ldots, \boldsymbol{X}_n$ are $n$
i.i.d. copies of $\boldsymbol{X}.$ For a given $\bbeta,$ let
$\hat{F}_{\bbeta,n}(x)$ denote the empirical c.d.f. of the corresponding
$n$ loss observations
$\ell(\bbeta^\intercal\boldsymbol{X}_1),\ldots,\ell(\bbeta^\intercal\boldsymbol{X}_n).$ Then
$v_\beta(\bbeta)$ can be estimated as
$\hat{v}_{\beta,n}(\bbeta) := \hat{F}_{\bbeta,n}^{-1}(1-\beta),$ which
is simply the $\lceil n(1-\beta) \rceil$-th order statistic from the
$n$ loss observations. The corresponding CVaR $C_\beta(\bbeta)$ can
then be estimated as, 
\begin{align}
  \hat{C}_{\beta,n}(\bbeta) := \hat{v}_{\beta,n}(\bbeta) +
  \frac{1}{n\beta} \sum_{i=1}^n \left[ \ell(\bbeta^\intercal\boldsymbol{X}_i)
  - \hat{v}_{\beta,n}(\bbeta)\right]^+,
  \label{eq:SA-CVaR}
\end{align}
which is simply the average over the highest $\lfloor n\beta \rfloor$
loss observations, \cite{HongVaR}. If the loss
$\ell(\bbeta^\intercal{\boldsymbol{X}})$ has finite variance, it
is well-known that this sample average estimator satisfies asymptotic
normality,
$\sqrt{n}(\hat{C}_{\beta,n}(\bbeta) - C_\beta(\bbeta))
\overset{\mathcal{L}}{\rightarrow} \sigma_{_{SA}}(\beta)
\mathcal{N}(0,1),$ where the limiting variance
$\lim_{n \rightarrow \infty} n \text{Var}[\hat{C}_{\beta,n}(\bbeta)]$
is given as in (see \cite{HongVaR}, Corollary 2),
\begin{align*}
  \sigma_{{\textnormal{SA}}}^2(\beta) :=
  \frac{1}{\beta^2} \text{Var}
  \left[ \big(\ell(\bbeta^\intercal\boldsymbol{X}) - v_{\beta}(\bbeta)\big)^+\right]. 
\end{align*}
Thus, if we aim to approximate the CVaR $C_\beta(\bbeta)$ within a
relative error of $\varepsilon,$ it is necessary that at least
$O(\beta^{-1}\delta^{-1}\varepsilon^{-2})$ samples of $\boldsymbol{X}$
are required to do so with $(1-\delta)$ confidence, 
\cite{HongCVaR}. Since this sample requirement is impractically large
when $\beta$ is small, importance sampling is employed, when feasible,
to reduce variance to a lower order than
$O(\beta^{-2}).$ 


\subsection{Importance sampling}
Importance sampling (IS) is a popular variance reduction technique
that has found applications in various engineering disciplines, most
notably in settings where rare events need to be tackled
\cite{Bucklew}. The objective of this section is to develop an
IS scheme for estimating $C_\beta(\bbeta)$ such that the resulting IS
estimator i) has substantially lower variance 
and ii) is applicable across nontrivial joint dependence among the
components of $\boldsymbol{X}.$ As we shall see in Section \ref{sec:DD},
the importance sampling estimator also serves as a natural starting
point towards estimating CVaR and its gradient from data when the
distribution of $\boldsymbol{X}$ is not known.

The first step in the IS estimation of CVaR is to develop an IS
estimator for the c.d.f.
$F_{\bbeta}(u) := 1 - P(\ell(\bbeta^\intercal\boldsymbol{X}) > u).$ Suppose that
the probability density of $\boldsymbol{X}$ is given by $f(\cdot).$ To
circumvent the issue of limited observations in the tail region
$E_1 := \{\boldsymbol{x}:\ell(\bbeta^\intercal\boldsymbol{x}) > u\}$ when $1-F_{\bbeta}(u)$
is small, we instead obtain samples for $\boldsymbol{X}$ from a carefully
chosen IS probability density $f_{_{IS}}(\cdot)$ under which the set
$E_1$ has much higher
probability than $1-F_{\bbeta}(u).$ Let
\[\mathcal{L}_R(\boldsymbol{x}) := \frac{f(\boldsymbol{x})}{f_{_{IS}}(\boldsymbol{x})}\]
denote the likelihood ratio associated with the change of
distribution. It is necessary that IS density is such that
$f_{_{IS}}(\boldsymbol{x}) > 0$ for any $\boldsymbol{x} \in E_1$ such that
$f(\boldsymbol{x}) > 0.$ Then,
\begin{align*}
  1-F_{\bbeta}(u) &= \int_{E_1} f(\boldsymbol{x})d\boldsymbol{x}
   = \int_{E_1} \frac{f(\boldsymbol{x})}{f_{_{IS}}(\boldsymbol{x})} f_{_{IS}}(\boldsymbol{x}) d\boldsymbol{x} \nonumber\\
  &= E\left[\mathcal{L}_R(\tilde{\boldsymbol{X}})\mathbb{I}(\ell(\bbeta^\intercal \tilde{\boldsymbol{X}}) > u)\right],
\end{align*}
where $\tilde{\boldsymbol{X}}$ is distributed according to
$f_{_{IS}}(\cdot).$ This suggests the use of the estimator,
\begin{align}
  \Fis(u) = 1 - \frac{1}{n}\sum_{i=1}^n \mathcal{L}_R\big(\tilde{\boldsymbol{X}}_i\big)
  \mathbb{I}\big(\ell(\bbeta^\intercal \tilde{\boldsymbol{X}}_i) > u \big),
  \label{eq:FIS}
\end{align}
where $\tilde{\boldsymbol{X}}_1,\ldots, \tilde{\boldsymbol{X}}_n$ are
i.i.d. samples from the IS density $f_{_{IS}}(\cdot).$ To obtain
estimators with low variance, it is desirable that the IS density is
such that the likelihood ratio
$\mathcal{L}_R(\tilde{\boldsymbol{X}}_i)$ is small. This is achieved
by choosing an IS density which somewhat mirrors the conditional
distribution of $\boldsymbol{X}$ over the target rare set $E_1$ (see
\cite{Bucklew}, Section 4.2). While there is a rich literature on the
choice of IS density when the components of $\boldsymbol{X}$ are
independent (see \cite{Lam} and references therein), the dependent
case is underdeveloped due to the need to carefully account for the
dependence structure present in the rare event of interest.

\subsection{The proposed IS density and the IS algorithm}
A central idea in extreme value theory is that when a distribution
satisfies Assumption \ref{assumption:GEV},
$\lim_{n \rightarrow \infty} n \Prob(\boldsymbol{X}/\boldsymbol{a}_n \in A)$
exists for a large class of sets $A,$ see \cite {Resnick}; as a
result, for suitably large $m,n$ such that $m < n,$
$\Prob(\boldsymbol{X}/\boldsymbol{a}_n \in A)$ can be approximated, in
principle, by $(m/n) \times \Prob(\boldsymbol{X}/\boldsymbol{a}_m \in A),$
thus indicating the use of self-similarity which is present at
different scales in the distribution of $\boldsymbol{X}.$

Exploiting this phenomenon towards the IS estimation of the c.d.f.
$F_{\bbeta}(u)$ results in the following radically new approach
towards selecing IS density: We first identify a lower level $l < u$
and suitably replicate the samples observed in the less rarer set
$E_{0} := \{\boldsymbol{x}: \ell(\bbeta^\intercal\boldsymbol{x}) > l\}$ by
appropriately scaling them onto the target rare set $E_1.$ To be
specific, if we take $\tilde{\boldsymbol{X}} = \boldsymbol{s} \boldsymbol{X}$ for
suitable scaling vector $\boldsymbol{s} = (s_1,\ldots,s_d)$  such that
$s_i \geq 1$ for all $i,$
then the resulting IS density is given by,
$f_{_{IS}}(\tilde{\boldsymbol{x}})$ = density of $\boldsymbol{s}\boldsymbol{X}$
$=f\left(\tilde{\boldsymbol{x}}/\boldsymbol{s}\right)\Pi_{k=1}^d s_k^{-1}.$
Consquently, the likelihood ratio is given by,
\begin{align}
  \mathcal{L}_R(\tilde{\boldsymbol{X}})
 &:= {f(\tilde{\boldsymbol{X}})}/{f_{_{IS}}(\tilde{\boldsymbol{X}})}\
   =  \left(\Pi_{k=1}^d s_k\right) f(\tilde{\boldsymbol{X}})/f(\boldsymbol{X}).
   \label{eq:LLR}
\end{align}
The self-similarilty of the distribution ensures that the IS density
mirrors the distribution of $\boldsymbol{X}$ over the set $E_1,$ thus
obviating the major impediment in searching for an IS density in
multivariate setting.  In other words, the proposed IS scheme
`automatically learns' the conditional distribution of $\boldsymbol{X}$
over the set $E_1$ by exploiting the more frequent samples observed in
a similar, but less rare, set $E_{0}.$

Since our interest in the c.d.f $F_{\bbeta}(\cdot)$ stems from the goal of
CVaR estimation, we present our IS scheme for jointly estimating
c.d.f., VaR and CVaR below. Variance reduction guarantees are
presented immediately after Algorithm \ref{algo:CVaR-I.S.}. Under
Assumption \ref{assumption:GEV}, it is necessary that the marginal
densities $f_k(\cdot)$ of $X_k, k = 1,\ldots,d$ and the joint pdf
$f(\cdot)$ of $\boldsymbol{X} = (X_1,\ldots,X_d)$ are polynomially decaying, as in,
\begin{align}
  \frac{f_k(tc)}{f_k(t)} \rightarrow c^{-(1+\alpha_k)} \text{ and } 
  \frac{f\big((ct)^{1/\boldsymbol{\alpha}}\boldsymbol{x}\big)}{f(t^{1/\boldsymbol{\alpha}}\boldsymbol{x})}
  \rightarrow c^{-(1+\sum_{k=1}^d \alpha_k^{-1})}
  \label{eq:MRV}
\end{align}
for any positive constant $c$ and
$\boldsymbol{x} \in \mathbb{R}_d^+ \setminus \{\boldsymbol{0}\},$ as
$t \rightarrow \infty$ (see \cite{Resnick}, Sections 5.2-5.5). Here,
$\boldsymbol{\alpha} = (\alpha_1,\ldots,\alpha_d)$ is the vector of
tail indices quantifying the rate of decay (or heaviness) of the
respective marginal tails of distributions. 
We have used
$t^{1/\boldsymbol{\alpha}} := (t^{1/\alpha_1},\ldots,t^{1/\alpha_d})$
to denote the component-wise exponentiation. 
Common examples of densities which satisfy \eqref{eq:MRV} include the
multivariate-t and Pareto densities (see \cite{deHaanRV}). 
\begin{algorithm}
 \caption{Importance Sampling Algorithm for computing CVaR}\label{algo:CVaR-I.S.}
  \textbf{Input:} $\beta_0$ such that the level
  $1-\beta_0 < 1- \beta$.\\
  \textbf{Procedure}\\
  \textbf{Step 1:} Generate $n$ i.i.d. samples
  $\boldsymbol{X}_1,\ldots,\boldsymbol{X}_n$ from $f(\cdot)$.\\
  \textbf{Step 2:} Let $s_k := (\beta_0/\beta)^{1/\alpha_k}$ for
  $k = 1,\ldots,d$ and $\boldsymbol{s} = (s_1,\ldots,s_d).$ Assign
  $\tilde{\boldsymbol{X}}_i = \boldsymbol{s}\boldsymbol{X}_i$ for
  $i = 1,\ldots,n.$ Compute the likelihood ratio from \eqref{eq:LLR}
  as
  \[\mathcal{L}_R(\tilde{\boldsymbol{X}}_{i}) = (\Pi_{k=1}^ds_k)
    f(\tilde{\boldsymbol{X}}_i)/f(\boldsymbol{X}_i).\]
\noindent \textbf{Step 3:} Compute the IS based \textnormal{CVaR} as,
   \begin{equation*}
       \cis (\bbeta) := \vis (\bbeta) + \frac{1}{n\beta}
      \sum_{i=1}^n \big(\ell(\bbeta^\intercal \tilde{\boldsymbol{X}}_i) - \vis (\bbeta) \big)^+
      \mathcal{L}_R(\tilde{\boldsymbol{X}}_i),
    \end{equation*}
    where IS based \textnormal{VaR},
    $\vis(\bbeta) := \inf\{ u : \Fis (u) \geq 1-\beta \},$ is
    estimated from the c.d.f. estimate $\Fis(\cdot)$ in
    \eqref{eq:FIS}.
  \end{algorithm}

  For the IS scheme in Algorithm \ref{algo:CVaR-I.S.}, we obtain the
  following variance guarantee as the CVaR estimation is made
  increasingly difficult by letting $\beta \searrow 0.$
  \begin{theorem}
    \label{thm:I-S-Var_Red}
    Suppose Assumptions~\ref{assume:l-loss} and \ref{assumption:GEV} hold, and
    $\textnormal{Var}[\ell(\bbeta^\intercal\boldsymbol{X})]$ is finite
    for $\bbeta \in \Theta.$ Further suppose that the joint
    probability density of $\boldsymbol{X},$ denoted by $f(\cdot),$ is
    approximated, as in (\ref{eq:MRV}), uniformly over
    $\boldsymbol{x} \in \mathbb{R}^d_+$ such that
    $\Vert \boldsymbol{x} \Vert = 1.$ Then the estimator
    $\cis(\bbeta)$ is asymptotically normal:
    \begin{align*}
      \sqrt{n}\big(\cis(\bbeta)  - C_\beta(\bbeta)\big) \overset{\mathcal{L}}{\rightarrow}
       \sigma_{{\textnormal{IS}}}(\beta)\mathcal{N}(0,1),
    \end{align*}
    where the limiting variance
    $\lim_{n \rightarrow \infty} n \text{Var}[\vis(\bbeta) ]$
    satisfies,
    \begin{align}
      \frac{\sigma_{{\textnormal{IS}}}^2(\beta)}{\sigma_{{\textnormal{SA}}}^2(\beta)}
      \leq \frac{\beta}{\beta_0} L_\sigma({\beta}/{\beta_0}) \left( 1 + o(1)\right),
      \label{eq:VR}
    \end{align}
    for a suitable slowly varying function $L_\sigma(\cdot),$ as
    $\beta_0,\beta$ are taken to 0.  In particular, the variance of the IS estimator, $\sigma_{\mathrm{IS}}^2(\beta)$, is vanishingly small relative to $\sigma_{\mathrm{SA}}^2(\beta)$, if we take $\beta_0 = \beta^k$, and the reduction
    guarantee \eqref{eq:VR} is uniform over $\bbeta \in \Theta.$
  \end{theorem}
  \begin{remark}
    \textnormal{To estimate CVaR $C_\beta(\bbeta)$ within a prescribed
      relative error,
      by taking $\beta_0 = \beta^\kappa$ for $\kappa < 1,$ we get
      variance reduction by a factor $\tilde{O}(\beta^{-(1-\kappa)}).$ Then,
      as a consequence of central limit theorem, it is sufficient to
      choose the number of samples for the IS scheme to be smaller by
      a factor
      $\tilde{O}(\beta^{-(1-\kappa)})$ 
      than naive sample averaging. 
    }
  \end{remark}

  \section{Extrapolation based data-driven estimators}
  \label{sec:DD}
  In this section, we consider the data-driven setting where the
  probability density of $\boldsymbol{X}$ is not known and the CVaR
  $C_\beta(\bbeta)$ and its gradient $\nabla C_\beta(\bbeta)$ have to
  be estimated from historical data
  $\boldsymbol{X}_1,\ldots,\boldsymbol{X}_n.$ 
  The starting point for the data-driven estimator is the IS CVaR
  estimator $\cis(\bbeta)$ in Step 3 of Algorithm
  \ref{algo:CVaR-I.S.}. Given i.i.d. data
  $\boldsymbol{X}_1,\ldots,\boldsymbol{X}_n,$ observe that all the
  terms in the expression for $\cis(\bbeta)$ are computable except the
  likelihood ratio $\mathcal{L}_R(\boldsymbol{s}\boldsymbol{X}).$ Due
  to the assumed heavy-tailed nature of $\boldsymbol{X},$ we can however
  approximate the likelihood ratio as demonstrated in Example
  \ref{eg:motivation}
  below. 

  \begin{example}
    \label{eg:motivation}
    \textnormal{
      Suppose that
      $L(\bbeta) = \bbeta^\intercal\boldsymbol{X}$ and the
      marginal components of $\boldsymbol{X} = ({X}_1,\ldots,{X}_d)$
      have the same distribution. Then $\alpha_k = \alpha$ for
      $k=1,\ldots,d.$ Given a tail probability level $\beta,$ take
      $\beta_0 = c\beta$ for a suitable constant $c > 1,$
      $t = 1/\beta_0$ and reparameterize
      $\boldsymbol{x} \in \mathbb{R}^d_{++}$ as
      ${\boldsymbol{z}} = t^{-1/{\alpha}}\boldsymbol{x}.$ To
      approximate the likelihood ratio
      $\mathcal{L}_R(s\boldsymbol{X}),$ recall that we have taken
      $s = c^{1/\alpha}$ in Algorithm \ref{algo:CVaR-I.S.}. Therefore,
      $s\boldsymbol{x} = (ct)^{{1/\alpha}}{\boldsymbol{z}}$ and
      $\boldsymbol{x} = t^{{1/\alpha}}{\boldsymbol{z}}.$ Since $t$ is
      large when the tail probability level $\beta$ is small, we can
      use (\ref{eq:MRV}) to approximate the likelihood ratio
      $\mathcal{L}_R(\cdot)$ as,
  \begin{align*}
    \mathcal{L}_R(s\boldsymbol{x}) =  c^{d/\alpha}
    \frac{f\big((ct)^{{1/\alpha}}{\boldsymbol{z}}\big)}
    {f\big(t^{{1/\alpha}}{\boldsymbol{z}}\big)}
    \approx c^{d/\alpha}c^{-(1+d/\alpha)} =  1/c, 
  \end{align*}
  for all $\boldsymbol{x}$ suitably large. In particular, since the
  target rare set
  $\{\boldsymbol{x}:\ell(\bbeta^\intercal\boldsymbol{x}) >
  v_{\beta}(\bbeta) \}$ can be shown to be contained in the set
  $\{\boldsymbol{x}: \Vert \boldsymbol{x} \Vert > \kappa
  t^{1/\alpha}\}$ for a suitable $\kappa$ (see Proposition \ref{lemma:CMT-WC}), the approximation
  $\mathcal{L}_R(\tilde{\boldsymbol{X}}_i) \approx 1/c =
  \beta/\beta_0$ is indeed applicable for all
  $\tilde{\boldsymbol{X}}_i = s\boldsymbol{X}_i$ for which
  $\ell(\bbeta^\intercal\tilde{\boldsymbol{X}}_i) > v_\beta(\bbeta)).$
  Then, from \eqref{eq:FIS}, we have the approximation,
  \begin{align*}
    1-\Fis(u) &\approx  \frac{1}{n}\sum_{i=1}^n\frac{\beta}{\beta_0} 
                \mathbb{I}\big( s\bbeta^\intercal \boldsymbol{X}_i > u \big)\\
              &= (\beta/\beta_0)
                \big(1-\hat{F}_{\bbeta,n}(u/s)\big),
  \end{align*}
  where
  $\hat{F}_{\bbeta,n}(x) := \frac{1}{n}\sum_{i=1}^n\mathbb{I}(\bbeta^\intercal
  \boldsymbol{X}_i \leq x)$ is the empirical c.d.f. constructed from
  the samples. 
  Recalling that
  $\ell(\bbeta^\intercal\boldsymbol{X}) = \bbeta^\intercal\boldsymbol{X},$ we obtain
  the following the expressions for $\vis(\bbeta),\cis(\bbeta)$ in
  Algorithm \ref{algo:CVaR-I.S.} and $\vsa,\csa$ in Section
  \ref{sec:SAA}: 
  \begin{align*}
    \vis(\bbeta)
    &\approx \inf\left\{u: 1- \frac{\beta}{\beta_0}
      \left[1-\hat{F}_{\bbeta,n}\left(\frac{u}{s}\right)\right] \geq 1-\beta \right\}\\
    &=  \inf\big\{su: 1- \hat{F}_{\bbeta,n}\left(u\right) \leq \beta_0\big\}
      = s\vsa(\bbeta), \text{ and }\\
    \cis (\bbeta) &\approx s\vsa (\bbeta) + \frac{1}{n\beta}
                    \sum_{i=1}^n \big(s\bbeta^\intercal {\boldsymbol{X}}_i - s\vsa (\bbeta) \big)^+
                    \frac{\beta}{\beta_0}\\
    &= s\csa(\bbeta) 
  \end{align*}
}
\end{example}

Theorem \ref{thm:convergence} below establishes that the above
heuristic approximations motivated by approximating the IS estimator
are indeed valid for the large class of losses $\ell(\cdot)$ and
diverse dependence structures for $\boldsymbol{X}$ introduced in the
beginning of Section \ref{sec:IS}.
\begin{theorem}
  \label{thm:convergence}
  Suppose Assumptions~\ref{assume:l-loss} and \ref{assumption:GEV} hold, and
  $\textnormal{Var}[\ell(\bbeta^\intercal\boldsymbol{X})]$ is finite
  for $\bbeta \in \Theta.$ Define
  $\xi := 1/\min_{k=1,\ldots,d}\alpha_k.$ Then,
  \begin{align}
    C_\beta(\bbeta) &= \left( \beta_0/\beta\right)^{\xi(\rho+1)} C_{\beta_0}(\bbeta)(1+o(1)) \quad \text{ and }\label{eqn:CVaR_Extrap-DD}\\
   \nabla C_\beta(\bbeta) &= \left( \beta_0/\beta\right)^{\xi(\rho+1)} \nabla C_{\beta_0}(\bbeta)(1+o(1))\label{eqn:CVaR_Extrap},
  \end{align}
  as $\beta \searrow 0$ and  $\beta_0/\beta \rightarrow c \in (1,\infty).$
\end{theorem}
The significance of the theorem is that the CVaR $C_\beta(\bbeta)$ and
its gradient $\nabla C_\beta(\bbeta)$ at a high probability level $1-\beta$
(say, $1-\beta = 0.99$) can be computed from their respective
estimates, $C_\beta(\bbeta)$ and $\nabla C_\beta(\bbeta),$ obtained from a
lower probability level $1-\beta_0$ (say, $1-\beta_0 = 0.9$). In
principle, this should reduce the number of samples required to
estimate the respective quantities by a factor of
$\beta_0/\beta.$ We observe this is indeed the case in most
numerical experiments in Section \ref{sec:num-exp}. While this type of
extrapolation performed from the risk observed at a lower level is the
central plank of tail risk measurement in extreme value theory in
statistics \cite{deHaan} and quantitative risk management
\cite{Mainik,mcneil}, to the best of our knowledge, such extrapolation
has not been utilized in gradient estimation and subsequent
optimization. The approximation for gradients
$\nabla C_\beta(\bbeta)$, as in Theorem \ref{thm:convergence}, is new
even from an extreme value theory point of view, and becomes an useful
addition to the arsenal of existing extrapolation techniques.
The approximations in Theorem \ref{thm:convergence} facilitates the
following low variance estimation scheme for the CVaR
$C_\beta(\bbeta)$ and its gradient $\nabla C_{\beta}(\bbeta).$
\begin{algorithm}[H]
   \label{algo:DD-Extrp}
    \caption{ Data Driven Algorithm for Evaluating CVaR and its Sensitivity}
\textbf{Input}: Samples $\boldsymbol{X}_1,\ldots,\boldsymbol{X}_n,$
   parameter $\bbeta,$ and levels $\beta,\beta_0.$\\
   \textbf{Procedure:}\\
\textbf{Step 1:} Compute $L_i = \ell(\bbeta^\intercal\boldsymbol{X}_i)$ and
   $L_i^\prime = \nabla \ell(\bbeta^\intercal\boldsymbol{X}_i),$ for $i=1,\ldots,n.$ \\
\textbf{Step 2: }Estimate VaR as
   $\hat{v}_{\beta_0,n}(\bbeta) = \lceil n(1-\beta_0) \rceil-$th order
   statistic of the collection $\{L_1,\ldots,L_n\}.$ Then estimate the
   CVaR and its gradient from the plug-in estimators below (see
   \cite{HongVaR,HongCVaR} respectively): 
   \begin{align}
     \hat{C}_{\beta_0,n}(\bbeta)
     &=  \hat{v}_{\beta_0,n}(\bbeta_0) + \frac{1}{n\beta_0} \sum_{i=1}^n
       \left[ L_i- \hat{v}_{\beta_0,n}(\bbeta)\right]^+,\nonumber\\
     \hat{\nabla}_n C_{\beta_0}(\bbeta)
     &= \frac{1}{n\beta_0} \sum_{i=1}^{n} L_i^\prime
       \mathbb{I}
       \big(L_i \geq \hat{v}_{\beta_0,n}(\bbeta)\big)\label{eqn:CVaR-DD}.
\end{align}
\textbf{Step 3: } Let $\{L_{n-i,n}\}_{i=0}^{n-1}$ denote the order statistics of the collection $\{L_1,\ldots,L_n\}$. Estimate the tail parameter $\hat\xi_n$ using the Hill estimator (see \cite{deHaan}, Section 3.2) as follows:
\begin{equation}\label{eqn:Hill-Est}
  \hat{\xi}_n = \frac{1}{n\beta_0}\sum_{i=0}^{\lfloor{n\beta_0}\rfloor-1} \log(L_{n-i,n}) - \log \left(L_{n-\lfloor{n\beta_0}\rfloor,n}\right).
\end{equation}
\textbf{Step 4: }Compute CVaR estimate $\tilde{C}_{\beta,n}(\bbeta)$ and its
gradient estimate $ \tilde{\nabla}_{n}C_\beta(\bbeta)$ at the target
level $\beta$ as follows:
\begin{align}\label{eqn:Extrp-DD}
  \tilde{C}_{\beta,n}(\bbeta) &= \hat{C}_{\beta_0,n}(\bbeta)(\beta_0/\beta)^{\hat{\xi}_n}\nonumber\\
  \tilde{\nabla}_{n}C_\beta(\bbeta) &= \hat{\nabla}_n{C}_{\beta_0,n}(\bbeta)(\beta_0/\beta)^{\hat{\xi}_n}.
\end{align}
Return $\tilde{C}_{\beta,n}(\bbeta),\tilde{\nabla}_{n}C_\beta(\bbeta).$
\end{algorithm}
Corollary~\ref{thm:Consistency} below establishes consistency of the
estimators (in a relative error sense) in
Algorithm~\ref{algo:DD-Extrp} as the target rare set
$\{L(\bbeta) > v_{\beta(n)}(\bbeta)\}$ is made increasingly rare; this
is accomplished by letting $\beta(n) \searrow 0$ as the number of
samples $n \rightarrow \infty.$

\begin{corollary}\label{thm:Consistency}
  Suppose $\beta(n)\searrow 0$ and
  $\beta_0(n)/\beta(n) \to c \in (1,\infty)$ as $n\to\infty.$ In
  addition, if the levels $\{\beta_0(n)\}_{n \geq 1}$ are chosen such
  that the respective sample average CVaR estimators
  $\{\hat{C}_{\beta_0}(n)\}_{n \geq 1}$ have vanishing relative error,
  then the estimators for target CVaR,
  $\{\tilde{C}_{\beta(n)}(\bbeta)\}_{n \geq 1}$ output by Algorithm
  \ref{algo:DD-Extrp}, also have vanishing relative error: that is,
  $ \tilde C_{\beta(n)} (\bbeta) - C_{\beta(n)}(\bbeta)
  =o_{\mathcal{{P}}}( C_{\beta(n)}(\bbeta)),$ as
  $n \rightarrow \infty.$
\end{corollary}
\begin{remark}
  \textnormal{
  The conditions of the Corollary~\ref{thm:Consistency} are satisfied,
  for example, when $n\beta_{0}(n) \rightarrow \infty.$ Similar
  consistency can be established for the CVaR gradient (by replacing
  the CVaR by its gradient everywhere in
  Corollary~\ref{thm:convergence}).}
\end{remark}
We now argue that the estimators in Algorithm~\ref{algo:DD-Extrp}
enjoy a variance reduction of $\beta_0/\beta$ over their naive
counterparts. To faciliatate this argument, recall that the error in
the IS estimator (see Section \ref{sec:IS}) for CVaR estimation is
approximately
$\mathcal{N}(0,n^{-1/2}\sigma^{2}_{IS}(\beta))$. Further, observe that
for the naive estimator of CVaR, $\hat{C}_\beta(\bbeta)$, from
Corollary 2 of \cite{HongVaR}, the error is roughly
$\mathcal{N}(0,n^{-1/2}\sigma^{2}_{SA}(\beta))$.  Recall that
Theorem~\ref{thm:I-S-Var_Red} shows that for small $\beta$,
$\sigma^{2}_{SA}(\beta)/\sigma^{2}_{IS}(\beta) \sim \beta_0/\beta$.
From Example~\ref{eg:motivation}, the estimators in
Algorithm~\ref{algo:DD-Extrp} were essentially arrived at by replacing
the likelihood in the importance sampling estimator from
Algorithm~\ref{algo:CVaR-I.S.} by its limit. Since the error of making
this approximation is negligible for small values of $\beta,$ the
proposed estimator $\tilde{C}_\beta(\bbeta)$ (and thus
$\tilde\nabla C_{\beta}(\bbeta)$) is likely to have variance smaller
by a factor $\beta_0/\beta$ than the naive sample average
counterparts.  In this case, the sample complexity becomes smaller by
a factor $\beta_0/\beta$ with the extrapolation scheme in
Algorithm~\ref{algo:DD-Extrp}. This is verified through the numerical
experiments in Section~\ref{sec:num-exp}.  Indeed, the above
reasoning may be made precise by deriving a central limit theorem for
the estimator from Algorithm~\ref{algo:DD-Extrp}, under the set-up of
Corollary~\ref{thm:convergence}, with some mild additional regularity
assumptions on the distribution of $\boldsymbol{X}$ (for e.g., the
second order conditions from \cite{deHaan}). However, this is beyond
the scope of the current paper, and will be pursued as a follow-up
research.
\section{Numerical Experiments}
\label{sec:num-exp}
In this section, we report the results of numerical experiments
performed with simulated and real data in order to compare the
performance of estimators proposed in Algorithm \ref{algo:DD-Extrp}
with that of the naive sample average estimators.
\subsection{Scalability of gradient estimation with dimension $d$}
\begin{figure*}[t!]
      \centering
\includegraphics[scale=0.4]{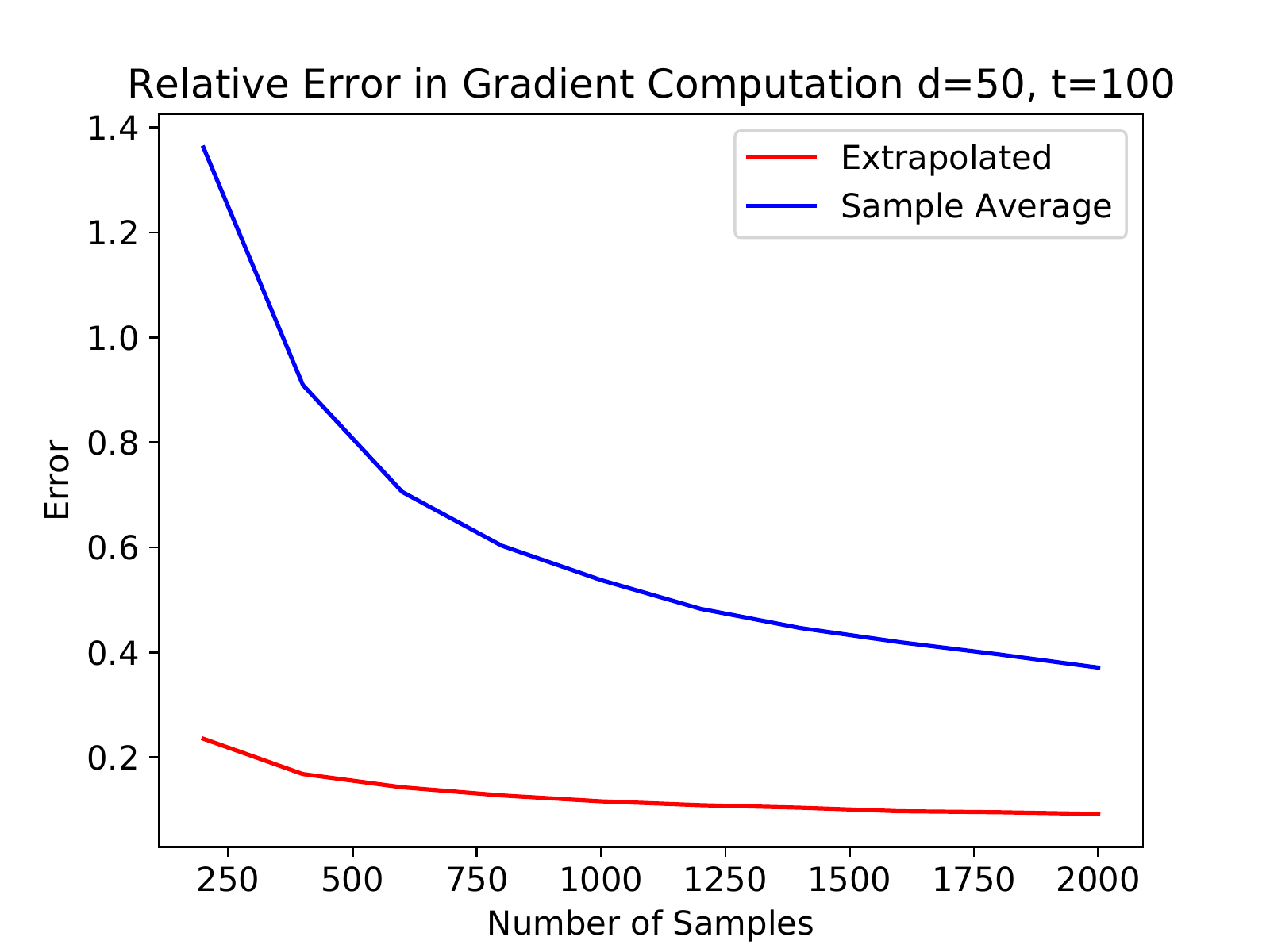}
\includegraphics[scale=0.4]{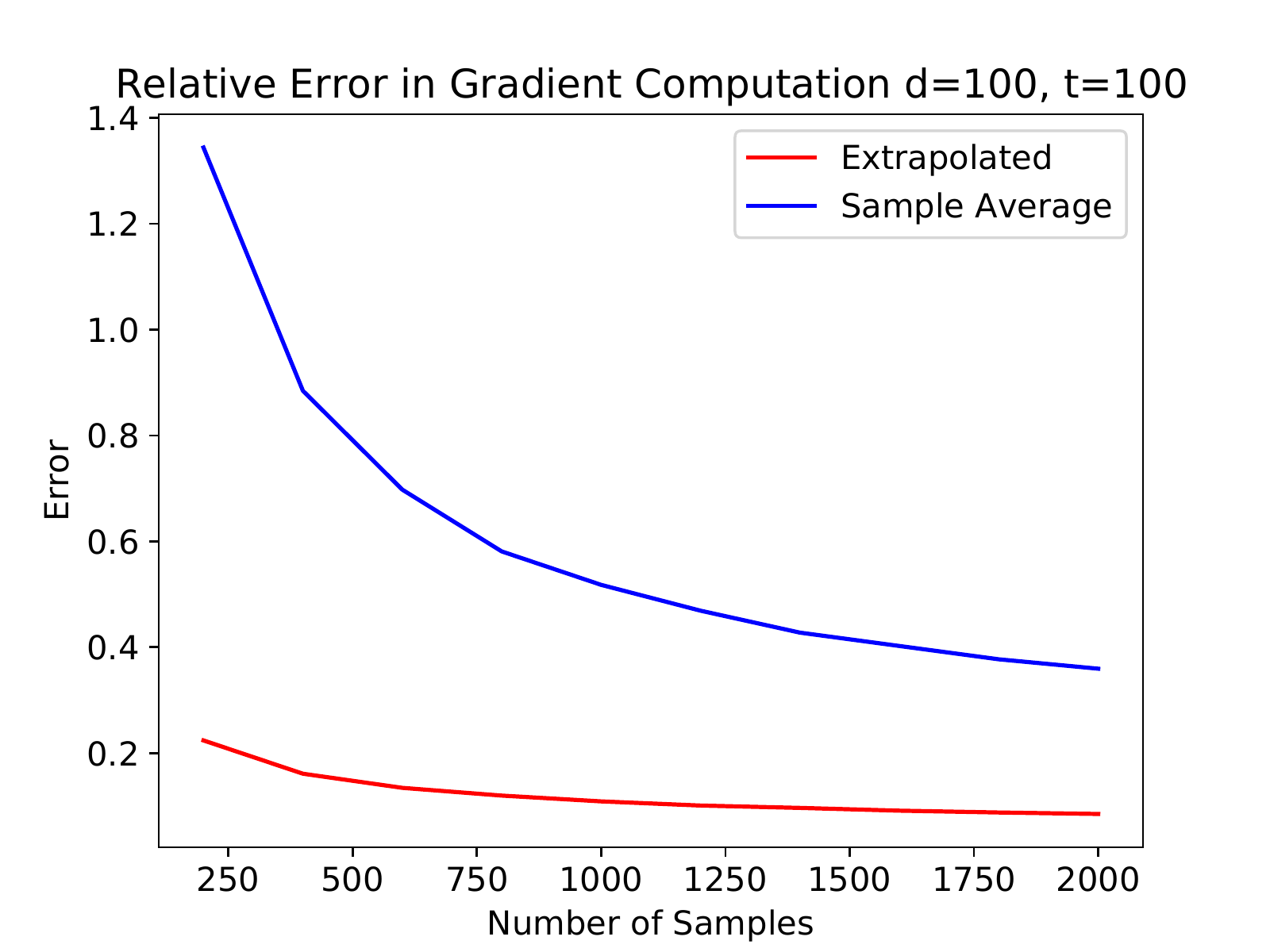}
\caption{Plots demonstrating the effectiveness of the extrapolation formula~\eqref{eqn:Extrp-DD} in Algorithm 3.1 when compared to naive sample average. The loss $L(\bbeta) = L_1(\bbeta)$, $d$ denotes the dimensionality, and $t=\beta^{-1}$ is the level at which the CVaR gradient is computed. The components of $\boldsymbol{X}$ are chosen to be independent Pareto(3)}\label{fig:Grad-Compute-Lin}
\end{figure*}

\begin{figure*}[h!]
 \centering
\includegraphics[scale=0.4]{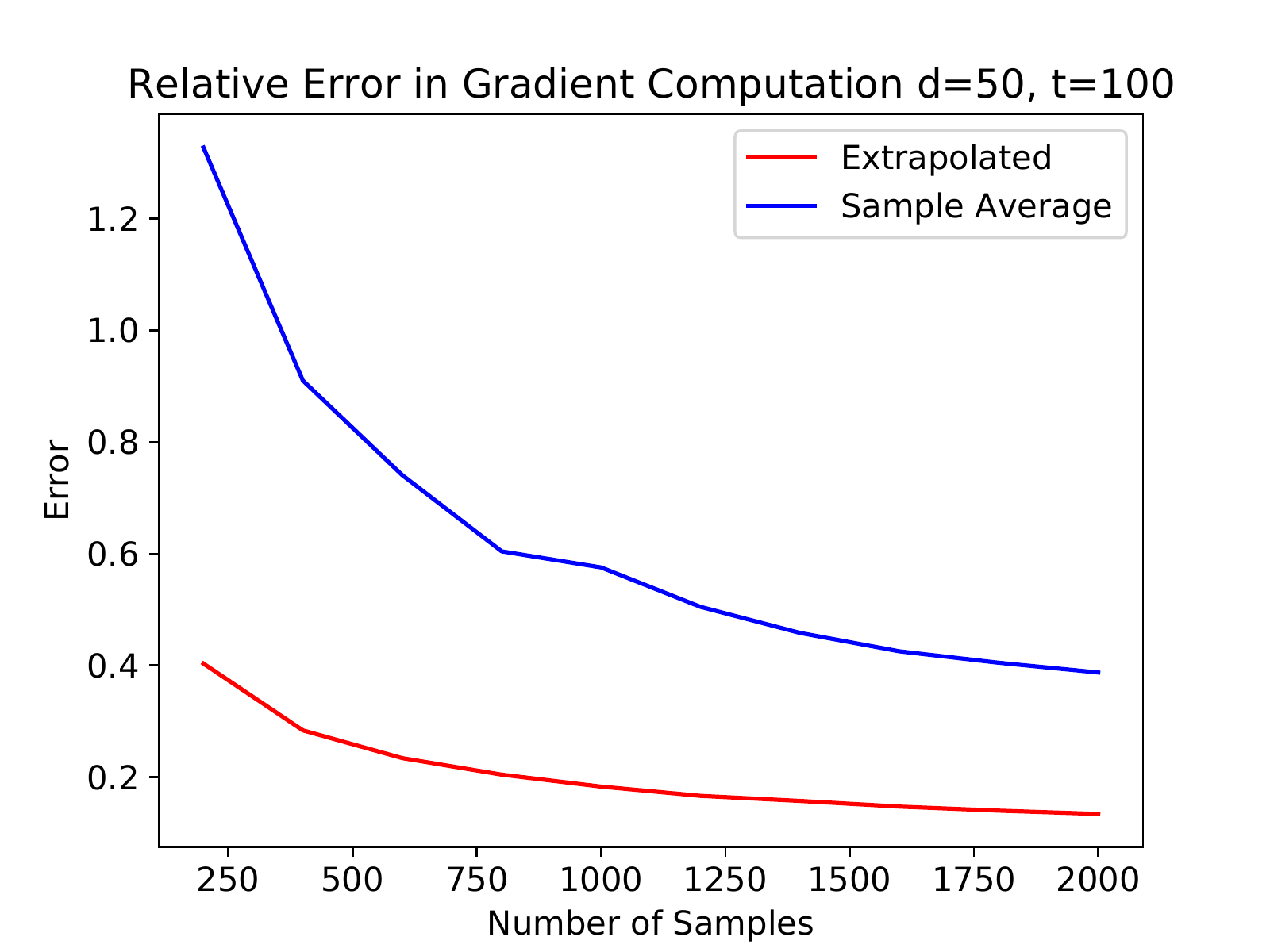}
\includegraphics[scale=0.4]{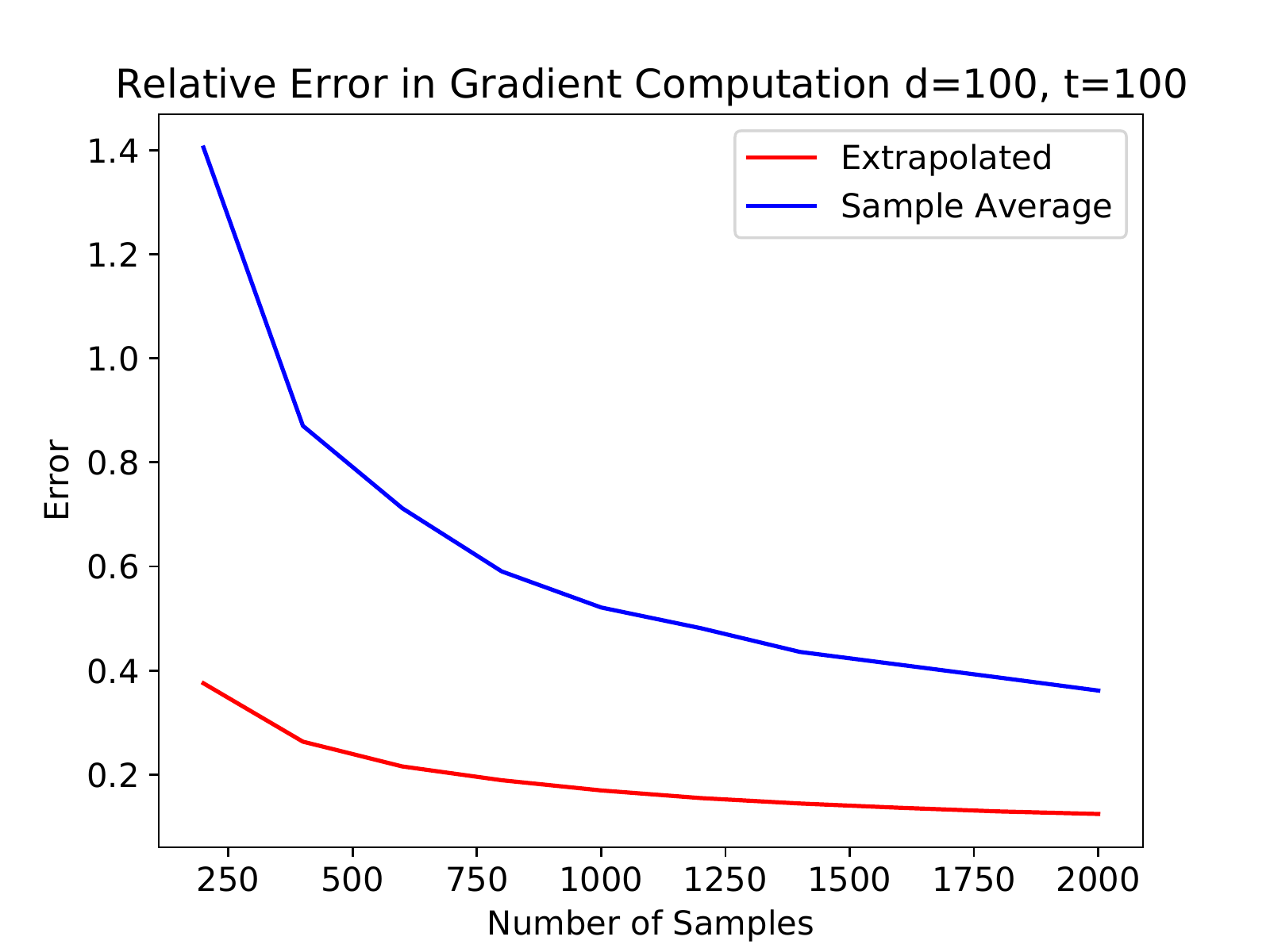}
\caption{The above plots show the efficacy of the estimator in \eqref{eqn:Extrp-DD} for computing the CVaR gradient. The loss function $L(\bbeta) = L_2(\bbeta)$, while the rest of the set up from Figure~\ref{fig:Grad-Compute-Lin} is used, but components of $\boldsymbol{X}$ are chosen to be independent Pareto(6).}\label{fig:Grad-Compute-sq}
\end{figure*}

In this experiment, we compare Root Mean Square Errors (RMSE) of the
relative errors of the gradients $\nabla C_{\beta}(\bbeta)$ computed
from i) the proposed estimator in Algorithm~\ref{algo:DD-Extrp} and
ii) the naive sample average (SA) estimator.  Taking $\beta = 0.1$ and
candidate losses to be $L_1(\bbeta) = \bbeta^\intercal \boldsymbol{X}$
and $L_2(\bbeta) = (\bbeta^\intercal \boldsymbol{X})^2,$ we report
RMSE relative errors observed in dimensions $d = 50$ and $d = 100.$
Since the true value of $\nabla C_{\beta}(\bbeta)$ are not known in
closed form, we approximate them using sample average with
$N_1 = 10^6 $ samples to serve as a benchmark for computing relative
error of our estimates.  For the linear loss $L_1(\bbeta),$ the lower
level $\beta_0$ is chosen as $20\beta$, whereas for the square loss
$L_2(\bbeta),$ $\beta_0$ is set at ${8}\beta$; these levels are
identified by cross-validating over an interval of candidate
$\beta_0$. The RMSE relative errors are reported in
Figures~\ref{fig:Grad-Compute-Lin} and \ref{fig:Grad-Compute-sq}.
It is evident from the figures that Algorithm~\ref{algo:DD-Extrp}
outperforms the naive method, robustly across dimensions $d = 50,100,$
by resulting in significantly smaller error than the naive method in
gradient estimation. Moreover, in a manner consistent with our
hypothesis, the number of samples required by the naive method to get
the same level of accuracy as the proposed estimator scales like
${\beta_0}/{\beta}:$ For example, in Figure~\ref{fig:Grad-Compute-sq},
where $d=50$ and $\beta=0.01$, fixing the RMSE observed for the
proposed estimator with $n = 250$ samples, we observe that the naive
sample estimator requires as much as $n=2000$ samples to offer the
identified RMSE.

\subsection{Application to Risk Constrained Portfolio Optimisation}
\begin{figure*}[t!]
     \centering
\includegraphics[scale=0.4]{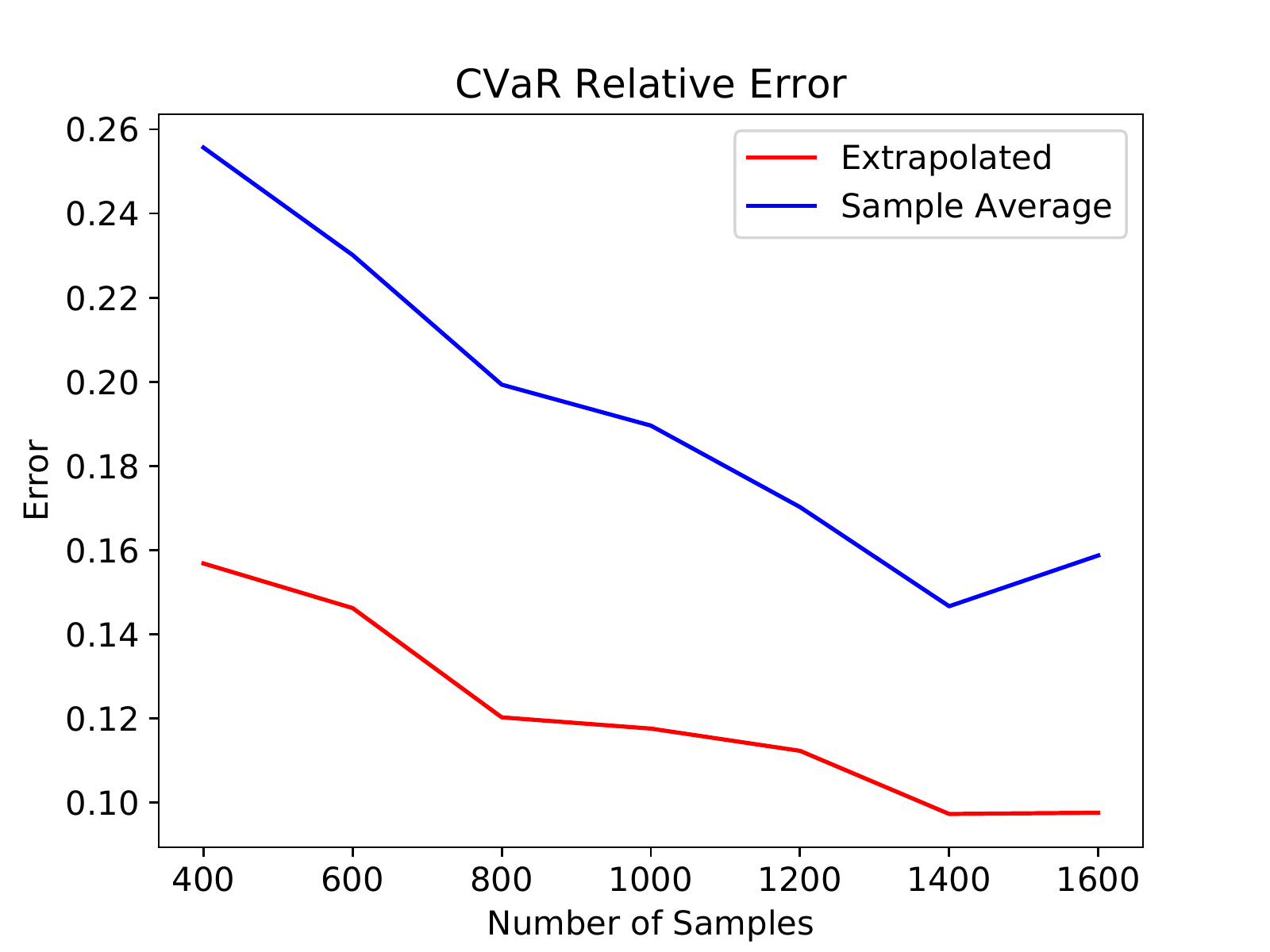}
\includegraphics[scale=0.4]{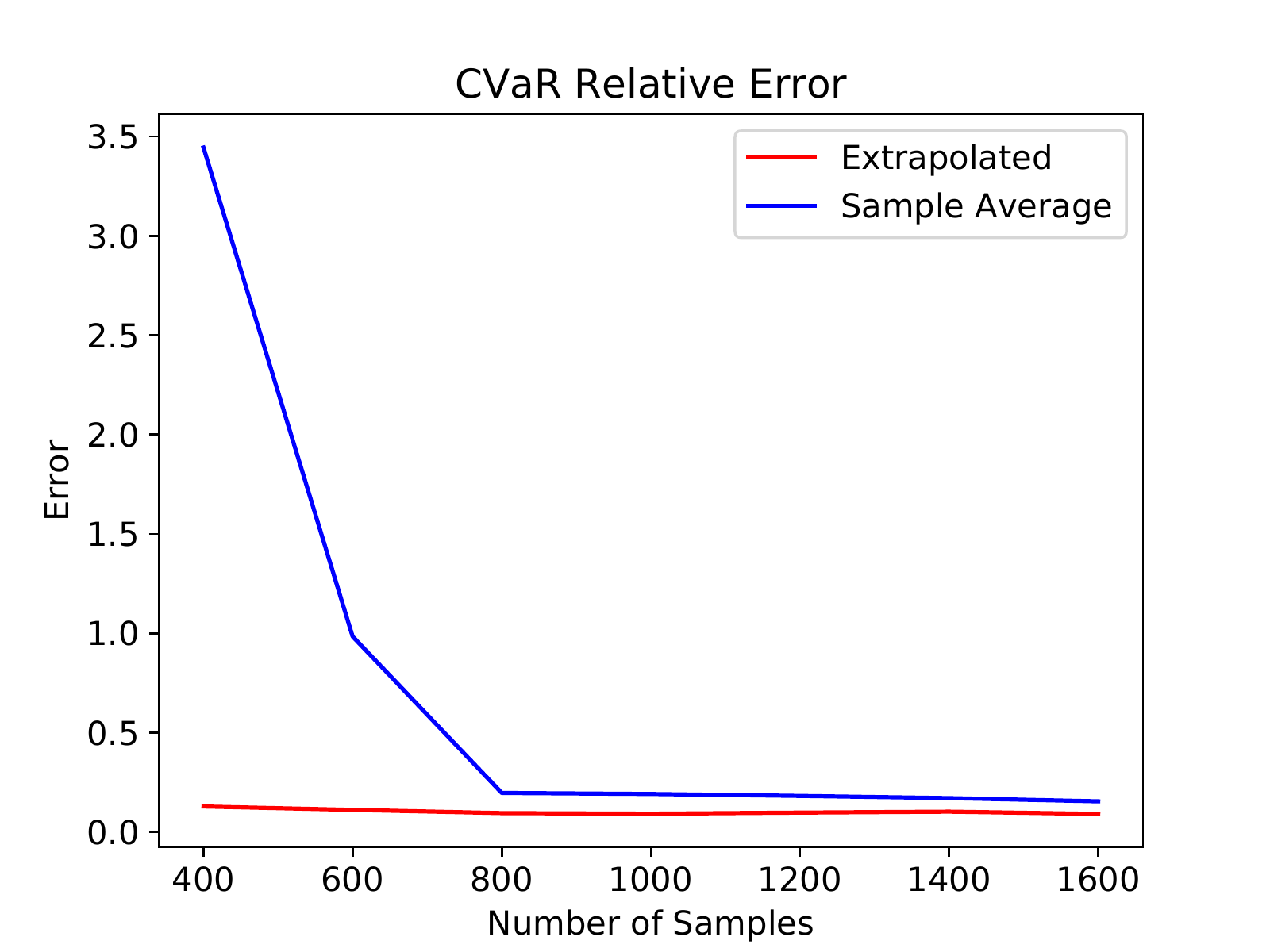}
\caption{The above plots demonstrate the effectiveness of Algorithm~\ref{algo:DD-Extrp} over sample average approximation, in solving the risk constrained problem  \eqref{eq:pf-opt}. Relative errors in the optimal CVaR in \eqref{eq:pf-opt} are compared. The panel  on the left indicates the errors when $\beta=0.01$, while  that on the right indicates the errors for $\beta=0.005$.} \label{fig:Grad-Compute-Opt}
\end{figure*}

We demonstrate the utility of the proposed gradient estimator by
applying it to solving a single period risk constrained portfolio
optimization problem.  The historical losses from $d$ assets are given
by observations $\{\boldsymbol{X}_1,\ldots, \boldsymbol{X}_n\}.$
Taking the loss $L(\bbeta)$ to be the portfolio loss
$\bbeta^\intercal\boldsymbol{X}$ for a given portfolio weight vector
$\bbeta,$ we aim to solve,
\begin{align}
  \min_{\bbeta \in  \mathbb{R}^d_+, \sum_i\theta_i =1} C_{\beta}(\bbeta).
  \label{eq:pf-opt}
\end{align}
This formulation has been considered extensively in the literature on
tail risk sensitive portfolio optimization (see \cite{Ban,Krokhmal,
  Lim}). In order to evaluate the effectiveness of the estimators, we
first compute the true optimum value $V^\ast$ of \eqref{eq:pf-opt}
using a large number ($N=10^5$) of samples to serve as a benchmark. We
then solve (\ref{eq:pf-opt}) using gradient descent, where the
gradients are estimated from limited data for both the proposed and
naive SA estimators. Suppose that $\bbeta_1$ and $\bbeta_2$ are the
optimal portfolio weights output by the gradient descent methods in
which the gradients are estimated, respectively, using the proposed
and the naive SA gradient estimators. We then evaluate the true
objective values $V_1,V_2$ at $\bbeta_1$ and $\bbeta_2$ by using
$N=10^5$ samples. We then report the relative mean square errors,
$\vert V_1 - V^\ast \vert/V^\ast$ indicating the efficacy of the
proposed gradient estimation scheme, and
$\vert V_1 - V^\ast \vert/V^\ast$ which indicates the efficacy of
the naive SA estimation scheme.\\
\textbf{Specific implementation and Results:} We assume that the
dependence between asset returns is captured using a $t$-copula and
allow the tails of losses from individual assets to be different. Such
an assumption is used widely in modelling asset returns (see
\cite{Glasserman}). In order to stay close to a realistic data set,
the marginal tails and dependence between asset returns are modelled
using the correlation matrix of daily returns from twelve S\&P-500
stocks, over a 5 year period (1200 days). We consider the cases where
$\beta=0.01, 0.005$ in \eqref{eq:pf-opt}. The number of times periods
of data $n$ is varied from $400$ days to $1600$ days (about 2-6
years). Increasing the amount of data beyond this is practically
infeasible, since typically only about 5-6 years of financial data is
used in portfolio optimisation problems (see \cite{Lim}). As before,
the lower level $\beta_0=0.1$ is selected through cross validation. We
observe that in Figure \ref{fig:Grad-Compute-Opt}, errors in the
optimal solution are significantly lower using when
Algorithm~\ref{algo:DD-Extrp} is used to compute CVaR gradients, over
using the naive estimator for CVaR gradient. For example, we observe
that the error in optimal solution using Algorithm~\ref{algo:DD-Extrp}
is roughly 17\%, when $\beta=0.005$ and $n=400$.  The number of
samples of data required to achieve the same level of accuracy using
the naive estimator is over 1600. This indicates a large reduction in
amount of data required to solve the risk constrained optimisation
problem (\ref{eq:pf-opt}) by means of the proposed method. 
\begin{figure*}[t!]
     \centering
\includegraphics[scale=0.4]{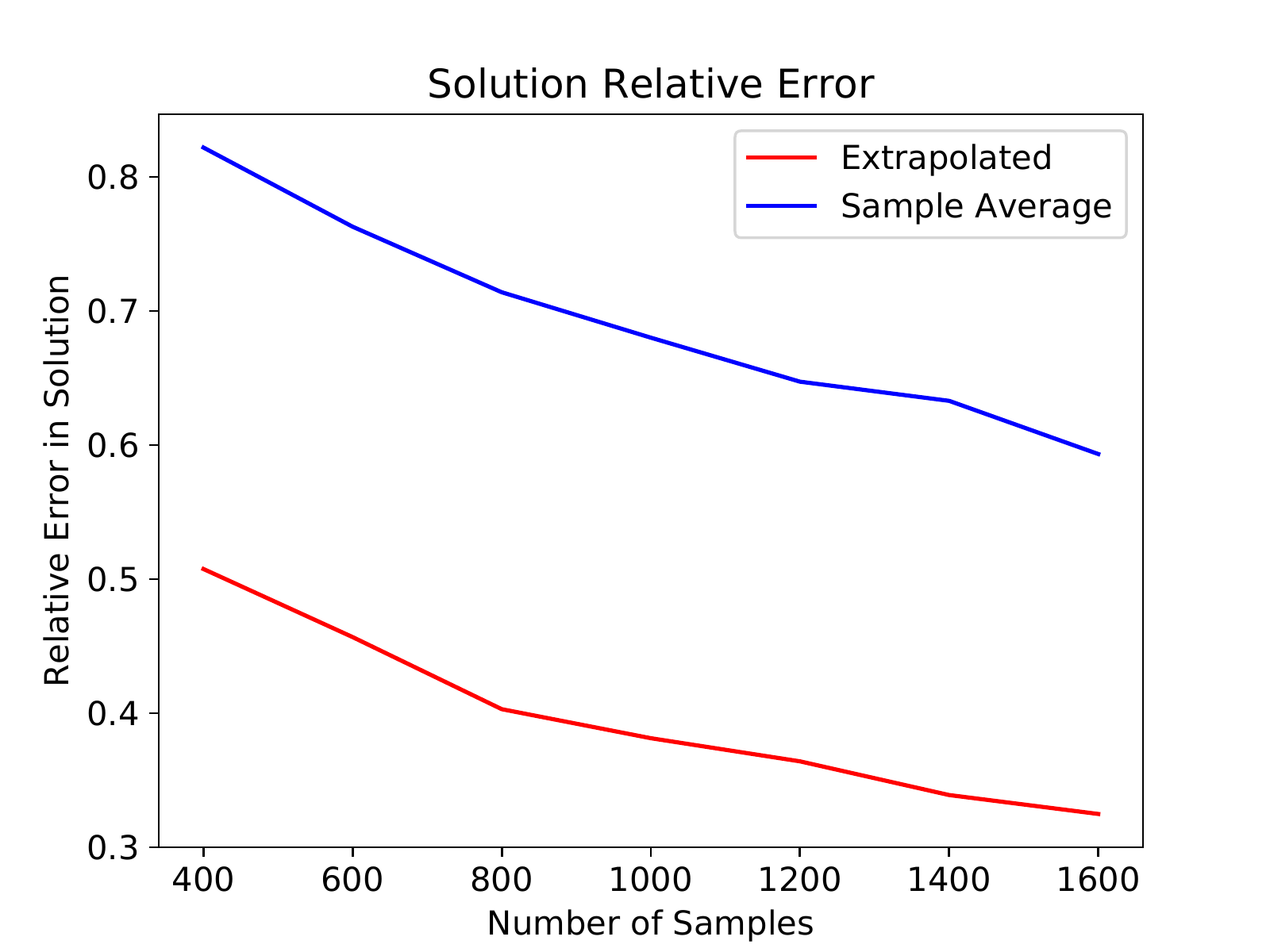}
\includegraphics[scale=0.4]{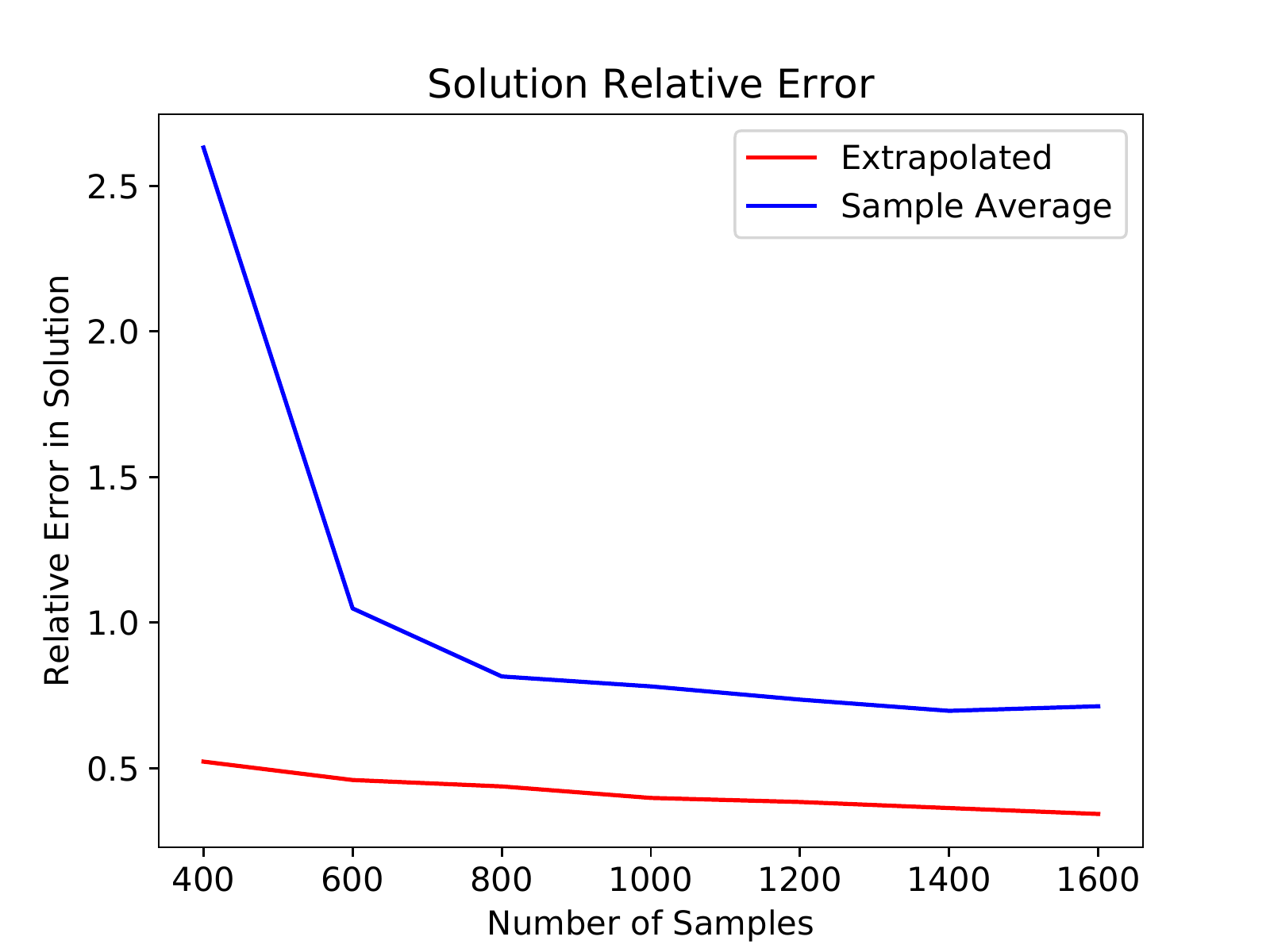}
\caption{The above plots demonstrate the effectiveness of Algorithm~\ref{algo:DD-Extrp} over sample average approximation, in solving the risk constrained problem  \eqref{eq:pf-opt}. Relative errors in the optimiser of \eqref{eq:pf-opt} are compared. The panel  on the left indicates the errors when $\beta=0.01$, while  that on the right indicates the errors for $\beta=0.005$.} \label{fig:Grad-Compute-Sol}
\end{figure*}
\subsection{Experiments with S\&P-500 data}
For demonstrating the efficacy of the proposed experiment with real
data, we use 5 years (1200 days) of daily returns for 12 S\&P-500
stocks. We divide the data into intervals, call them $S_i$, of the
form $[N_i,N_i+300]_{i=1}^{100}$. Here, the starting indices $N_i$ are
chosen uniformly at random, without replacement from
$\{1,\ldots, 1200\}$. For each sample window $S_i$, we compute the
CVaR and its gradient using the naive approach and the proposed
Algorithm~\ref{algo:DD-Extrp} for the choices $\beta_0=0.1$ and
$\beta=0.01$. We then calculate the mean and the variance of both the
estimators. We find that for the CVaR, the variance of naive
computation is $303.15$, while that for the extrapolation based
estimator is $161.77$ (the means in computation of CVaR are 41.05 and
41.64, respectively). Similarly, for the gradient, the variance for
the naive SA estimator is $403.74$, while that for the proposed
extrapolation based estimator is $163.88$. This suggests that even on
a representative sample of real asset data, the proposed estimator
significantly out-performs the naive sample average.

 \section{Proofs of Theorem~ \ref{thm:I-S-Var_Red} and \ref{thm:convergence}}  \label{sec:proofs}
We present the proofs for the case where $\ell(u)=u$.  We begin by establishing an asymptotic approximation for $v_{\beta}(\bbeta)$.  Notice that under Assumption~\ref{assumption:GEV}, from  Theorem 3 of \cite{Rootzen} and Theorem 5.3 of \cite{Resnick},  as $t\uparrow\infty$.
\begin{equation}\label{eqn:Resnick-PP}
t\Prob\left({\boldsymbol{X}}/{\mathbf{a}_t} \in \cdot\right) \to \nu(\boldsymbol{z}\in\cdot ),    
\end{equation}
where the convergence is over all sets not containing the origin, and $\nu(\cdot)$ is a Radon measure (see \cite{Rootzen} for a detailed explanation). Thus, the  tail behaviour of $\boldsymbol{X}$ can be inferred by looking at the dependence measure $\nu(\cdot)$. Specifically, for a suitably large $t$, for any $\boldsymbol{b}\in \Real^d_+$, $\Prob(\boldsymbol{b}^{\intercal}\boldsymbol{X}/\mathbf{a}_t \geq u) \approx t^{-1} \nu(\boldsymbol{b}^{\intercal}\mathbf{z} \geq u)$. Define $\bar{a}_t=\|\boldsymbol{{a}}_t\|_{\infty}$, let $\bar{\mathbf{\bbeta}} = \lim_{t \rightarrow \infty}\bar{a}_t^{-1}\mathbf{a}_t\bbeta$ and $\hat{\mathbf{a}} := \lim_t \bar{a}_t^{-1}\mathbf{a}_t$. Further define $\boldsymbol{X}_t=\boldsymbol{X}/\mathbf{a}_t$. Then, under Assumption~\ref{assumption:GEV}, the limits $\bar{\bbeta}$ and $\hat{\mathbf{a}}$ are well defined. Observe that with $\bbeta_t  = \bar{a}_t^{-1}\mathbf{a}_t\bbeta$, one can write $\bar{a}_t^{-1}\bbeta^\intercal\boldsymbol{X} = \bbeta_t^\intercal\boldsymbol{X}_t$. Proposition~\ref{lemma:CMT-WC} extends the discussed convergence slightly:
\begin{proposition}\label{lemma:CMT-WC}
\begin{equation}
    \lim_{t\to\infty}t\Prob\left( \bbeta^\intercal \boldsymbol{X} \geq \bar{a}_t u\right) = \nu(\bar\bbeta^\intercal\boldsymbol{z} \geq u).
\end{equation}
\end{proposition} 
For any $s \leq 1,$ define 
\begin{align}
    \kappa_s({\bbeta}) & := \inf\left\{u: \nu\left(\bar{\bbeta}^T\boldsymbol{z} > u \right) \leq s \right\}\label{defn:kappa-0}.
\end{align}
Proposition~\ref{lemma:CMT-WC} shows that for a large $t$, the probability that $\bbeta^\intercal\boldsymbol{X}\geq \bar{a_t}u$ is approximately $t^{-1}\nu(\bar\bbeta^\intercal\boldsymbol{z} \geq u)$. By setting $\nu(\bar\bbeta^\intercal\boldsymbol{z} \geq u)$ to 1, $v_{\beta}(\bbeta) \sim  \bar{a}_t \kappa_1(\bbeta)$,  as $t \rightarrow \infty$ (see Proposition~\ref{prop:Var-Convex}). The proof of Proposition~\ref{lemma:CMT-WC}, which outlines how the limiting measure $\nu(\cdot)$ plays a vital role in all our analysis, is given in Section~\ref{sec:Proof-1}.
\subsection{Proof of Theorem \ref{thm:convergence}}\label{sec:Proof-thm2}
We present the proof of Theorem~\ref{thm:convergence}. Proofs of the intermediate technical steps are given in Appendix~\ref{sec:Proofs}. Henceforth, throughout this proof, let $t=1/\beta$, and $t_0=1/\beta_0$. Steps 1 and 2 aim at establishing the following approximation:
\begin{equation}\label{eqn:ApproxDer}
    \nabla C_{\beta}(\bbeta) = \bar{a}_t(g(\bbeta))(1+o(1)).
\end{equation}
\noindent \textit{\textbf{Step 1:} }The first step is to approximate the VaR of $\bbeta^\intercal\boldsymbol{X}$ using Proposition~\ref{lemma:CMT-WC}.  Proposition~\ref{prop:Var-Convex}, below gives the necessary approximation:
\begin{proposition}\label{prop:Var-Convex}
Under the assumptions of Theorem \ref{thm:convergence}, $v_{\beta}(\bbeta) \sim  \bar{a}_t \kappa_1(\bbeta)$,  as $t \rightarrow \infty$.
\end{proposition}
From Proposition~\ref{prop:Var-Convex}, $v_{\beta}(\bbeta) = (1+\epsilon_t)\bar{a}_t\kappa_1(\bbeta)$, where $\epsilon_t\to 0$. Now, for all $t$, 
\[
C_{\beta}(\bbeta) = \bar{a}_t\Expc\left( \bbeta^{\intercal}{\boldsymbol{X}}/{\bar{a}_t} \,\big \vert \bbeta^\intercal \boldsymbol{X}\geq \bar{a}_t(1+\epsilon_t) \kappa_1(\bbeta)\right),
\]
and from Theorem 1 of \cite{HongCVaR},  $\nabla C_{\beta}(\bbeta)$ equals
\begin{align}
    & \nabla \bar{a}_t \left( \Expc\left({\bbeta^\intercal\boldsymbol{X}}/{\bar{a}_t} \, \big \vert \,  \bbeta^\intercal \boldsymbol{X}\geq \bar{a}_t(1+\epsilon_t) \kappa_1(\bbeta)\right)\right) \nonumber\\
    &= \bar{a}_t\Expc\left({\boldsymbol{X}}/{\bar{a}_t} \, \big \vert \, \bbeta^\intercal \boldsymbol{X}\geq \bar{a}_t(1+\epsilon_t) \kappa_1(\bbeta)\right).\label{eqn:Interchange-der}
\end{align} 
Proposition~\ref{prop:Conditional-Lim} below aids in approximating the conditional expectation in \eqref{eqn:Interchange-der}
\begin{proposition}\label{prop:Conditional-Lim}
    Let $\bar\bbeta_t\to\bar\bbeta$ as $t\to\infty$. Then,
\begin{equation}
        \mathcal{L}\left({\boldsymbol{X}}/{\bar a_t} \, \big \vert \, \bar\bbeta_t^\intercal\boldsymbol{X}_t \geq  \kappa_1(\bbeta) \right) \xrightarrow{\mathcal{L}} \mu,
    \end{equation}
where for any $B\in\Real^d$, $ \mu(B) = \nu(\hat{\mathbf{a}}\boldsymbol{z}\in B\, , \bar\bbeta^\intercal\boldsymbol{z}\geq \kappa_{1}(\bbeta))$.
\end{proposition}
Let  $\bar\bbeta_t=\frac{\bbeta\mathbf{a}_t}{{\bar{a}}_t(1+\epsilon_t)}$. Then, observe that $$\nabla C_{\beta}(\bbeta) = \bar{a}_t \Expc\left({\boldsymbol{X}}/{\bar{a}_t} \, \big \vert \,  \bar\bbeta_t^\intercal \boldsymbol{X}_t\geq  \kappa_1(\bbeta)\right),  $$ and that $\bar{\bbeta}_t\to\bar{\bbeta}$. Therefore, from Proposition~\ref{prop:Conditional-Lim} the conditional law of $ \boldsymbol{X}$ given $\bbeta^\intercal \boldsymbol{X} \geq v_{\beta}(\bbeta)$ converges to $\mu(\cdot)$. \\
\noindent \textit{\textbf{Step 2:} }
In order to establish convergence of expectations from the above weak convergence, the following uniform integrebility condition is needed:
 \begin{lemma}\label{lemma:UI}
For any sequence $\bar\bbeta_t\to\bar\bbeta$, ${\boldsymbol{X}}/{\bar{a}_t}$ conditioned on $\bar\bbeta_t^\intercal \boldsymbol{X}_t \geq \kappa_1(\bbeta)$ is uniformly integreble. 
\end{lemma}
 Then, from Proposition ~\ref{prop:Conditional-Lim}, Lemma~\ref{lemma:UI}, and Theorem 3.5 of \cite{Billingsley}, $\Expc\left({\boldsymbol{X}}/{\bar{a}_t} \, \big \vert \, \bbeta^\intercal \boldsymbol{X}\geq v_t(\bbeta)\right)$ equals
\[
\Expc\left({\boldsymbol{X}}/{\bar{a}_t} \, \big \vert \, \bbeta^\intercal \boldsymbol{X}\geq \bar{a}_t(1+\epsilon_t) \kappa_1(\bbeta)\right) \to \Expc(\boldsymbol{Y}),
\]
where $\boldsymbol{Y}$ is distributed as
\begin{equation}\label{eqn:mu-law}
\mu(d\boldsymbol{y}) =  \nu(\hat{\mathbf{a}}\boldsymbol{z} \in d\boldsymbol{y}, \bar\bbeta^\intercal\boldsymbol{z} \geq\kappa_1(\bbeta)).
\end{equation}
Notice that since $\nu(\bar{\bbeta}^{\intercal} \boldsymbol{z} \geq \kappa_1(\bbeta)) =1$, $\mu(\Real^d) =1$, and it is a probability measure (see the proof of Lemma~\ref{lemma:FinConv} for a precise expression for $\mu(\cdot)$). Writing $g(\bbeta) = E\boldsymbol{Y}$,
\[
\Expc\left({\boldsymbol{X}}/{\bar{a}_t} \, \big \vert \, \bbeta^\intercal \boldsymbol{X}\geq \bar{a}_t(1+\epsilon_t)\right) -g(\bbeta) = o(1).
\]
This establishes \eqref{eqn:ApproxDer}.\\
\textit{\textbf{Step 3:} }Since $\beta\to 0$, we can write
\[
\nabla C_\beta(\bbeta)= \bar{a}_{t}g(\bbeta)(1+o(1)) = \bar{a}_t/ \bar{a}_{t_0} \times \bar{a}_{t_0}g(\bbeta)(1+o(1)).
\]
\noindent  Recall that $\xi=\max_{i=1}^{d} \xi_i$ is the index of the slowest decaying tail among $(X_1,\ldots,X_{d})$, and that under Assumption~\ref{assumption:GEV}, for every $i$, $a_t^i$ is regularly varying with rate $\xi_i$. Then, $\bar{a}_t$ is regularly varying with rate $\xi$. From Theorem 4.3.8 of \cite{deHaan} for all $t <t_0/K_1$ (or $\beta >K_1 \beta_0$) $\bar{a}_{t_0}\left( {t}/{t_0}\right)^{\xi}= \bar{a}_{t} (1+o(1))$.
Thus, $\nabla C_\beta(\bbeta)= \bar{a}_{t} g(\bbeta)(1+o(1))$ equals 
\[
 \left( {t}/{t_0}\right)^{\xi} \bar{a}_{t_0} g(\bbeta)(1+o(1)).
\]
Recall that  $t=1/\beta$ and $t_0=1/\beta_0$. Applying the approximation in \eqref{eqn:ApproxDer} (but with $\beta_0$ instead) gives \eqref{eqn:CVaR_Extrap}. To get \eqref{eqn:CVaR_Extrap-DD}, notice that applying the continuous mapping theorem (see \cite{Billingsley}, Theorem 2.7) to the weak convergence in Proposition~\ref{prop:Conditional-Lim}, with $f(\xx) = \bbeta^\intercal \xx$, the convergence
\[
\mathrm{Law}\left(\bbeta^\intercal \boldsymbol{X}/a_t \vert \ \bbeta^\intercal \boldsymbol{X} \geq v_t(\bbeta) \right) \to \mu_1
\]
is obtained, where $\mu_1$ is the push-forward measure associated with the map $\boldsymbol{z} \mapsto f(\boldsymbol{z})$, and $\mu_1(dy) =\nu(\bar{\bbeta}^\intercal\boldsymbol{z} \in dy; \bar{\bbeta}^\intercal \boldsymbol{z} \geq \kappa_1(\bbeta))$. Following uniform integrebility in Lemma~\ref{lemma:UI}, 
\[
C_{t}(\bbeta) = \bar{a}_t g_1(\bbeta)(1+o(1)).
\]
Now, repeating Step 3, with $\nabla C_t$ replaced by $C_t$ establishes \eqref{eqn:CVaR_Extrap-DD}.\qed\\
As mentioned earlier, to demonstrate the role of the limiting measure $\nu(\cdot)$ in the analysis, we present the proof of Proposition \ref{lemma:CMT-WC}. 
\subsection{Proof of Proposition \ref{lemma:CMT-WC}:}\label{sec:Proof-1}
Recall that the limits $\bar\bbeta$ and $\boldsymbol{\hat{a}}$ are well defined. Hence, for all large enough $t$, $\|\Delta_t\| = \|\bar\bbeta-\bbeta_t\| \leq \delta$. Writing $\bbeta_t=\bbeta+\Delta_t$, from the Cauchy-Schwarz inequality, for all large enough $t$,
\begin{align*}
    \Prob(\bar\bbeta^\intercal \boldsymbol{X}_t  -  \delta\|\boldsymbol{X}_t\| > u)\leq \Prob(\bbeta_t^\intercal\boldsymbol{X}_t \geq u)\leq \Prob(\bar\bbeta^\intercal \boldsymbol{X}_t  +  \delta\|\boldsymbol{X}_t\| \geq u).
\end{align*}
Observe that for $u>0$,
\[
\mathbf{0} \notin \left\{ \boldsymbol{z}: \bar\bbeta^\intercal\boldsymbol{z}-\delta\|\boldsymbol{z}\|>u\right\} \cup \left\{ \boldsymbol{z}: \bar\bbeta^\intercal\boldsymbol{z}+\delta\|\boldsymbol{z}\|\geq u\right\}.
\]
Now, by the Portmanteau Theorem for vague convergence (see Proposition 3.12 of \cite{Resnickbook}), 
\[
\liminf_{t\to\infty} t\Prob(\bar\bbeta^\intercal \boldsymbol{X}_t  -  \delta\|\boldsymbol{X}_t\| > u)\geq \nu (\bar\bbeta^\intercal\boldsymbol{z}-\delta\|\boldsymbol{z}\|>u)
\]
and
\[
\limsup_{t\to\infty} t\Prob(\bar\bbeta^\intercal \boldsymbol{X}_t  +  \delta\|\boldsymbol{X}_t\| > u)\leq \nu (\bar\bbeta^\intercal\boldsymbol{z}+\delta\|\boldsymbol{z}\|\geq u)
\]
Define $A_{n} = \left\{  \boldsymbol{z}: \bar\bbeta^\intercal\boldsymbol{z} -\|\boldsymbol{z}\|/n >u \right\}$  and  $B_{n} = \left\{  \boldsymbol{z}: \bar\bbeta^\intercal\boldsymbol{z} +\|\boldsymbol{z}\|/n\geq u \right\}$.
Then,
\[
A_n\uparrow \left\{ \boldsymbol{z}: \bbeta^\intercal\boldsymbol{z} >u\right\} \textrm{\, and \,} B_n\downarrow \left\{ \boldsymbol{z}: \bbeta^\intercal\boldsymbol{z} \geq u\right\}
\]
Since $\nu(\cdot)$ is a Radon measure, it assigns finite mass to compact sets. Since $\mathbf{0}\not\in B_1$, $\nu(B_n) < \infty$ for all $n$. By the continuity of measure, it follows that $\nu(A_n) \to \nu(\boldsymbol{z}:\bar\bbeta^\intercal\boldsymbol{z} > u)$ and  $\nu(B_n) \to \nu(\boldsymbol{z}:\bar\bbeta^\intercal\boldsymbol{z} \geq u)$ as $n\to\infty$
Now, for $\delta$ small enough, 
\[
\nu(\boldsymbol{z}:\bar\bbeta^\intercal\boldsymbol{z} > u) - \nu(\boldsymbol{z}:\bar\bbeta^\intercal\boldsymbol{z}-\delta \|\boldsymbol{z}\| > u) \leq \epsilon,
\]
and hence, for all $\epsilon>0$,
\[
\liminf_{t\to\infty} t\Prob(\bbeta_t^\intercal \boldsymbol{X}_t \geq u)\geq \nu (\bar\bbeta^\intercal\boldsymbol{z}>u) -\epsilon.
\]
\begin{lemma}\label{lemma:Zero-mass}
For any $u>0$, $\nu(\bar\bbeta^\intercal\boldsymbol{z}=u)=0$.
\end{lemma}
\noindent From Lemma~\ref{lemma:Zero-mass}, since $\epsilon$ was arbitrary,
\[
\liminf_{t\to\infty} t\Prob(\bar\bbeta_t^\intercal \boldsymbol{X}_t  \geq u)\geq \nu (\bar\bbeta^\intercal\boldsymbol{z} \geq u) 
\]
Similarly, $\limsup_{t\to\infty} t\Prob(\bar\bbeta_t^\intercal \boldsymbol{X}_t \geq u)\leq \nu (\bar\bbeta^\intercal\boldsymbol{z} \geq u)$,  which completes the proof.
\qed\\
\subsection{Proof of Theorem~\ref{thm:I-S-Var_Red}} For simplicity, we only demonstrate the proof for the case $l(u)=u$. As before, let $\beta = 1/t$ and $\beta_0=1/t_0$. Let $\mathcal{L}(\yy)$ be the likelihood ratio. The central limit theorem follows directly as a consequence of Corollary 2 of \cite{HongVaR}. Further,  $\sigma_{{\textnormal{IS}}}^2(\beta) = t^2 (\VarT((\bbeta^\intercal\boldsymbol{Y} - v_{\beta})^{+}\mathcal{L}(\boldsymbol{Y}) ) $,
where $\VarT$ denotes the variance with respect to the alternate measure. This equals
\[
\tilde{\Expc} \left(  (\bbeta^\intercal \boldsymbol{X} - v_{\beta})^{2}\mathbf{I}(\bbeta^\intercal\boldsymbol{X} \geq v_{\beta})\mathcal{L}^2(\boldsymbol{X}) \right) \textrm{ minus }
\]
\[
{\Expc}^2 \left(  (\bbeta^\intercal \boldsymbol{X} - v_{\beta})^{2}\mathbf{I}(\bbeta^\intercal\boldsymbol{X} \geq v_{\beta}) \right).
\]
In order to prove the second part of the theorem, notice that following a change of measure
\[
\tilde{\Expc} \left(  (\bbeta^\intercal \boldsymbol{X} - v_{\beta})^{2}\mathbf{I}(\bbeta^\intercal\boldsymbol{X} \geq v_{\beta})\mathcal{L}^2(\boldsymbol{X}) \right) \textrm{ equals }\] 
\begin{equation}\label{eqn:IS-Transf}
\Expc\left(  (\bbeta^\intercal \boldsymbol{X} - v_{\beta})^{2}\mathbf{I}(\bbeta^\intercal\boldsymbol{X} \geq v_{\beta})\mathcal{L}(\boldsymbol{X}) \right).
\end{equation}
Further, recall that from the Jacobian formula, $\mathcal{L}(\boldsymbol{X})$ is 
\[
\prod_{i=1}^{d}\left( {t}/{t_0}\right)^{\frac{1}{\alpha_i}} {f(\boldsymbol{X})}/{f\left(\boldsymbol{X}\left({t_0}/{t}\right)^{{1}/{\alpha}}\right)}.
\]
Now, \eqref{eqn:IS-Transf} becomes $\prod_{i=1}^{d}\left(t/t_0\right)^{{1}/{\alpha_i}}$ times
\begin{equation}\label{eqn:IS-Transf-2}
\int_{\bbeta^\intercal\boldsymbol{x} \geq v_{\beta}} (\bbeta^\intercal\boldsymbol{x} -v_{\beta})^2  {f(\xx)}/{f\left(\boldsymbol{x}\left({t_0}/{t}\right)^{{1}/{\alpha}}\right)} f(\xx) d\xx.
\end{equation}
A few definitions are needed to proceed. Let $\theta_{i,t} = u_t^{\alpha^{-1}_i - \alpha_*^{-1}} \theta_i $, where $u_t= v_{\beta}^{\alpha_*}$. Now, let $\boldsymbol{p} = u_t^{-{1}/{\alpha}}\boldsymbol{x}$. Then, \eqref{eqn:IS-Transf-2} becomes $\prod_{i=1}^{d}\left( {t}/{t_0}\right)^{{1}/{\alpha_i}} v_{\beta}^{2}$ times $\prod_{i=1}^d u_t^{1/\alpha_i}$ times
\[
\int_{\bbeta_t^\intercal\boldsymbol{p} \geq 1} (\bbeta_t^\intercal\boldsymbol{p} -1)^2  \frac{f\left( u_t^{1/\alpha}\boldsymbol{p} \right)}{f\left(\boldsymbol{p}\left({u_t\cdot t_0}/{t}\right)^{1/\alpha}\right)} f({u_t^{1/\alpha}\boldsymbol{p}}) d\boldsymbol{p}
\] 
Notice that for all large enough $t$,  $\inf_{\bbeta_t^\intercal\boldsymbol{p} \geq 1}\|\boldsymbol{p}\| \geq \delta$, for some $\delta>0$. Fix $\epsilon>0$.  From the hypothesis of the theorem (see \cite{Resnick}, Section 6, p.g. 199 onward), for all large enough $t$, uniformly over $\boldsymbol{p}\not\in B_{\delta}^{c}$, 
\[
f(u_t^{1/\alpha} \boldsymbol{p}) \leq (1+\epsilon)L\left(u_t\right) u_t^{-1} \prod_{i=1}^d u_t^{-1/\alpha_i}\varphi(\boldsymbol{p}), 
\]
and $f\left(\boldsymbol{p}\left(\frac{u_t\cdot t_0}{t}\right)^{1/\alpha}\right)$ is at least
\[
(1+\epsilon)^{-1} L\left({u_t\cdot t_0}/{t}\right)\left({u_t\cdot t_0}/{t}\right)^{-1} \prod_{i=1}^{d} \left({u_t\cdot t_0}/{t}\right)^{-1/\alpha_i} \varphi(\boldsymbol{p}),
\]
where $L(\cdot)$ is a slowly varying function. Thus, \eqref{eqn:IS-Transf-2} is upper bounded  by $(1+\epsilon)^{3}$ times $t/t_0$ times
\begin{equation}\label{eqn:6}
v_{\beta}^{2-\alpha_*} L^{2}\left(u_t\right)/L\left({u_t\cdot t_0}/{t}\right) \int_{\bbeta_t^\intercal \boldsymbol{p}  \geq 1} (\bbeta_t^\intercal\boldsymbol{p} -1)^2 \varphi(\boldsymbol{p}) d\boldsymbol{p}.
\end{equation}
Finally, observe that as $t\to\infty$, $\int_{\bbeta_t^\intercal \boldsymbol{p}  \geq 1} (\bbeta_t^\intercal\boldsymbol{p} -1)^2 \varphi(\boldsymbol{p}) d\boldsymbol{p} $ is  $(1+o(1))$ times $\int_{\bbeta_\infty^\intercal \boldsymbol{p}  \geq 1} (\bbeta_\infty^\intercal\boldsymbol{p} -1)^2 \varphi(\boldsymbol{p}) d\boldsymbol{p}$.
Define 
\[
\kappa(\bbeta) = \int_{\bbeta_\infty^\intercal \boldsymbol{p}  \geq 1} (\bbeta_\infty^\intercal\boldsymbol{p} -1)^2 \varphi(\boldsymbol{p}) d\boldsymbol{p},\] 
where $\bbeta_\infty =\lim_t\bbeta_t$.  To conclude, 
\begin{equation}\label{eqn:Lim_Small}
\Expc^2\left((\bbeta^\intercal\boldsymbol{X}  - v_\beta)^{+} \right)=   o\left(\Expc\left( ( \bbeta^\intercal\boldsymbol{X}  - v_\beta)^{+} \right)^2\right).
\end{equation}
Thus, to analyse asymptotic variance of the naive estimator, it is sufficient to analyse the first term above, call it $I_1$. Further, note that if $u\to\infty$, 
\begin{equation}\label{eqn:Lim_Expc}
\Expc\left((\bbeta^\intercal\boldsymbol{X}  - u)^{+} \right)^2 \sim L(u^{\alpha_*})u^{2-\alpha_*}\kappa(\bbeta).
\end{equation}
Putting $u=v_{\beta}$ completes the first part of the proof.  Plugging in $t=1/\beta$ and $t=1/\beta_0$, with $\beta_0=\beta^\kappa$ for all sufficiently small $\beta$,  since $L(\cdot)$ is slowly varying, from Karamata's representation (see \cite{deHaan}, Proposition B.1.6), ${\sigma_{{\textnormal{IS}}}^2(\beta)}/{\sigma_{{\textnormal{SA}}}^2(\beta)}\leq \beta^{1-\kappa-\epsilon}$, for $\epsilon>0$ arbitrary. This gives the exponential variance reduction. To  see that this variance reduction is uniform over $\bbeta\in \Theta$, observe that the statements below \eqref{eqn:IS-Transf-2} hold whenever for all $\{\boldsymbol{p}:\bbeta^\intercal\boldsymbol{p}\geq 1\}$,  $\|\boldsymbol{p}\|\geq \delta$. Since $\bbeta$ lies in a compact set not containing the origin, $\|\bbeta\|<M_K<\infty$ for all $\theta\in \Theta$. Thus, the previous statement holds uniformly, for all $\bbeta\in\Theta$. This implies that the approximations following \eqref{eqn:IS-Transf-2}, hold uniformly, and thus, so does the variance reduction.\qed\\
\section{Conclusions}
In this paper, we develop extrapolation based estimators for the computation of CVaR and its derivative. Such extreme value based estimators for CVaR and its gradient are entirely new, and have not been studied in literature. We apply this estimator to a risk constrained portfolio optimisation problem, and find that the extrapolation based methods achieve a substantial improvement in performance over the naive approach. Broadly, extreme value based extrapolations should have a wide range of applications in finance and operations research, where problems involving risk constraint optimisation with limited data are common.
\section{Acknowldegements}
The authors gratefully acknowledge the support of the Department of
Atomic Energy, Government of India (under project
no. 12-R\&D-TFR-5.01-0500) and the Singapore Ministry of Education
(under project MOE SRG 2018 134).

\appendix
\section{Statements and Proofs of intermediate Lemmas}\label{sec:Proofs}
Recall that $\bar{a}_t = \|\boldsymbol{a}_t\|_\infty$, $\hat{\boldsymbol{a}}_t = \boldsymbol{a}_t/\bar{a}_t$, $\boldsymbol{X}_t = \boldsymbol{X}/\boldsymbol{a}_t$, ${\bbeta}_t= \bbeta \hat{\boldsymbol{a}}_t$, $\hat{\boldsymbol{a}} = \lim_{t\to\infty} \hat{\boldsymbol{a}}_t$, and $\bar{\bbeta} = \hat{\boldsymbol{a}}\bbeta$.

\textbf{Proof of Proposition~\ref{prop:Var-Convex}} Recall that $v_t(\bbeta) = \inf_{u}\Big\{ \Prob\left( \bbeta^\intercal \boldsymbol{X} \geq u\right) = \frac{1}{t}\Big\}$. Observe that for all $u$, we can write $\Prob\left( \bbeta ^{\intercal} \left(\frac{\boldsymbol{X}}{\bar{a}_t}\right) \geq u\right) = \Prob\left( \bbeta ^{\intercal}_{t} \left(\frac{\boldsymbol{X}}{\mathbf{a}_t}\right) \geq u\right)$. Recall that from Proposition~\ref{prop:Var-Convex}, for all $u>0$, $tP(\bbeta_t^\intercal \boldsymbol{X}_t \geq u) \to \nu(\bbeta^{\intercal}\boldsymbol{z} \geq u)$. Observe that $ \nu(\bbeta^{\intercal}\boldsymbol{z} \geq \kappa_1(\bbeta)) =1$. Then, setting $u= \kappa_1(\bbeta)$, for all $t$ large,
\[
\vert t\Prob\left( \bbeta ^{\intercal}{\boldsymbol{X}}  \geq \bar{a}_t\kappa_1(\bbeta) \right) - 1\vert \leq \epsilon.
\]
Then with $\bar{F}^{-1}_{\bbeta}(\cdot)$ as the inverse complimentary CDF of $\bbeta^\intercal\boldsymbol{X}$, for all $t$ sufficiently large, 
\[
\bar{F}^{-1}_{\bbeta}\left(\frac{(1+\epsilon)}{t}\right) \leq \bar{a}_t\kappa_1(\bbeta) \leq \bar{F}^{-1}_{\bbeta}\left(\frac{(1-\epsilon)}{t}\right)
\]
Observe that since $\bar{a}_t$ is regularly varying with rate $\xi$, so is $\bbeta^\intercal\boldsymbol{X}$. Using Potter's bounds (see Proposition B.1.9 of \cite{deHaan}) $\bar{F}^{-1}_{\bbeta}\left(\frac{(1-\epsilon)}{t}\right)$ is bounded above by
\[
\bar{F}^{-1}_{\bbeta}\left(t^{-1}\right)\left((1+\epsilon)^{\xi} +\epsilon (1+\epsilon)^{\xi-1}\right), 
\]
and $\bar{F}^{-1}_{\bbeta}\left(\frac{(1+\epsilon)}{t}\right)$  is bounded below by
\[ \bar{F}^{-1}_{\bbeta}\left(t^{-1}\right)\left((1-\epsilon)^{\xi} -\epsilon (1-\epsilon)^{\xi+1}\right).\]
Recall that since $\boldsymbol{X}$ was non-atomic, so is $\bbeta^\intercal \boldsymbol{X}$. Thus, $F^{-1}_{\bbeta}(t^{-1}) = v_t(\bbeta)$. Then, 
\begin{align*}
&\left((1-\epsilon)^{\xi} -\epsilon (1-\epsilon)^{\xi-1}\right) \leq \frac{\bar{a}_t\kappa_1(\bbeta)}{v_{t}(\bbeta)}\\
&\leq \left((1+\epsilon)^{\xi} +\epsilon (1+\epsilon)^{\xi+1}\right).    
\end{align*}
Since $\epsilon$ was arbitrary, the proof is complete.\qed

\textbf{Proof of Proposition~\ref{prop:Conditional-Lim}} In order to prove Proposition~\ref{prop:Conditional-Lim}, we show that for all $\boldsymbol{X}\in\Real^d$ such that $\nu(\hat{\mathbf{a}}\boldsymbol{z} =\boldsymbol{x} , \bar\bbeta^\intercal\boldsymbol{z}\geq \kappa_{1}(\bbeta))=0$, $\Prob\left( {\boldsymbol{X}}/{\bar{a}_t} \leq \boldsymbol{x} \,\big\vert \, \bbeta_t^\intercal\boldsymbol{X}_t\geq  \kappa_1(\bbeta)\right)$ converges to $\nu(\hat{\mathbf{a}}\boldsymbol{z} \leq\boldsymbol{x} , \bar\bbeta^\intercal\boldsymbol{z}\geq \kappa_{1}(\bbeta))$.
To this end, observe that $\Prob\left( {\boldsymbol{X}}/{\bar{a}_t} \leq \boldsymbol{x} \,\big\vert \, \bbeta_t^\intercal\boldsymbol{X}_t \geq  \kappa_1(\bbeta)\right)$ equals
\begin{align}
\frac{t\Prob\left( {\boldsymbol{X}}/{\bar{a}_t} \leq \boldsymbol{x} , \, \bbeta_t^\intercal\boldsymbol{X}_t \geq  \kappa_1(\bbeta)\right) }{t\Prob\left( \bbeta_t^\intercal\boldsymbol{X}_t \geq  \kappa_1(\bbeta)\right) }\label{eqn:CDP-ratio}
\end{align}
The denominator of \eqref{eqn:CDP-ratio} converges to 1 by definition of $\kappa_1(\bbeta)$ (see the proof of Proposition~\ref{prop:Var-Convex}). Now, defining $\mathbf{\hat{a}}_t = {\mathbf{a}_t}/{\bar{a}_t}$,
\begin{align}\label{eqn:Prob-Common}
    &\Prob\left( {\boldsymbol{X}}/{\bar{a}_t} \leq \boldsymbol{x} , \, \bbeta_t^\intercal\boldsymbol{X}_t \geq  \kappa_1(\bbeta)\right)\nonumber\\
    &= \Prob\left(\hat{\mathbf{a}}_t \boldsymbol{X}_t \leq \boldsymbol{x} , \, \bbeta_t^\intercal\boldsymbol{X}_t \geq  \kappa_1(\bbeta)\right)
\end{align}
Since $\hat{\mathbf{a}}_t$ converges to $\hat{\mathbf{a}}$ as $t\to \infty$, $\hat{\mathbf{a}}_t = \hat{\mathbf{a}} +\ddelta_t$ for some $\ddelta_t\to\mathbf{0}$. Fix a vector $\ddelta=(\delta_1\ldots,\delta_d)$. Then, for a sufficiently large $t$, component-wise, $-\ddelta\leq\ddelta_t\leq\ddelta$. For any vector $\boldsymbol{X}\in\Real^d$, write $\vert\boldsymbol{X}\vert$ for the vector with components $(\vert x_1\vert,\ldots, \vert x_d \vert)$. Now, for all sufficiently large $t$, $\Prob\left(\hat{\mathbf{a}}_t \boldsymbol{X}_t \leq \boldsymbol{x} , \, \bbeta_t^\intercal\boldsymbol{X}_t 
\geq  \kappa_1(\bbeta)\right)$ is bounded below by $\Prob\left((\hat{\mathbf{a} } + \vert\ddelta\vert)\boldsymbol{X}_t < \boldsymbol{x} , \, \bar\bbeta^\intercal\boldsymbol{X}_t  - \delta_1 \|\boldsymbol{X}_t\|>  \kappa_1(\bbeta)\right)$. By the Portmanteau theorem, for all $\ddelta$, $\liminf_{t\to \infty} t\Prob\left( {\boldsymbol{X}}/{\bar{a}_t} \leq \boldsymbol{x} , \, \bbeta_t^\intercal\boldsymbol{X}_t \geq  \kappa_1(\bbeta)\right)$ is bounded below by $  \nu( (\mathbf{\hat{a}} + |\ddelta|) \boldsymbol{z} < \boldsymbol{x}, \, \bar\bbeta^\intercal\boldsymbol{z}  - \delta_1 \|\boldsymbol{z}\|>  \kappa_1(\bbeta) )$.  Define the sets $F_n$ as
\[
\left\{\boldsymbol{z} : \left(\mathbf{\hat{a}} + \big\vert{\mathbf{1}}/{\mathbf{n}}\big\vert\right) \boldsymbol{z} < \boldsymbol{x}, \, \bar\bbeta^\intercal\boldsymbol{z}  - {1}/{n}\|\boldsymbol{z}\|>  \kappa_1(\bbeta)  \right\}
\]
Observe that $F_n$ are open and
\[
F_n\uparrow F=\left\{ \boldsymbol{z} : \mathbf{\hat{a}} \boldsymbol{z} < \boldsymbol{x}, \, \bar\bbeta^\intercal\boldsymbol{z} >  \kappa_1(\bbeta) \right\}
\]
Now, using the continuity of measure, since $\ddelta$ above was arbitrary, for all $\epsilon>0$
\begin{align*}
&\nu( (\mathbf{\hat{a}} + |\ddelta|) \boldsymbol{z} < \boldsymbol{x}, \, \bar\bbeta^\intercal\boldsymbol{z}  - \delta_1 \|\boldsymbol{z}\|>  \kappa_1(\bbeta) )\\
&\geq \nu( \mathbf{\hat{a}}\boldsymbol{z} \leq \boldsymbol{x}, \, \bar\bbeta^\intercal\boldsymbol{z} \geq  \kappa_1(\bbeta) ) -\epsilon \text{ since $\nu(\mathbf{\hat{a}}\boldsymbol{z}  = \xx , \bar{\bbeta}^\intercal \boldsymbol{z} = \kappa_1(\bbeta)) =0$, by assumption},
\end{align*}
that is, $\liminf_{t\to \infty} t\Prob\left( {\boldsymbol{X}}/{\bar{a}_t} \leq \boldsymbol{x} , \, \bbeta_t^\intercal\boldsymbol{X}_t \geq  \kappa_1(\bbeta)\right)$ is lower bounded by 
$\nu( \mathbf{\hat{a}}\boldsymbol{z} \leq \boldsymbol{x}, \, \bar\bbeta^\intercal\boldsymbol{z} \geq  \kappa_1(\bbeta) )$. Let $I=\{i\in[d]: \hat{a}_i =0\}$. Now, for any $\ddelta =(\delta_1,\ldots,\delta_d)$ define the new vector $\hat{\mathbf{a}}_{\ddelta}$ as
\begin{align*}
    (\hat{\mathbf{a}_{\ddelta}})_i &= 0 \textrm{ if $i\in I$ }\\
    &= \hat{a}_i -\delta_i \textrm{ otherwise}.
\end{align*}
Observe that since $\hat{\mathbf{a}}_t \to \hat{\mathbf{a}}$,  and $\bbeta_t\to\bar{\bbeta}$, for all sufficiently large $t$, $\hat{a}^i_t\geq (\hat{a}^i -\delta)\vee 0$. Thus, for all large enough $t$,
\begin{align*}
\left\{\hat{\mathbf{a}}_{\ddelta}\boldsymbol{X}_t\leq \boldsymbol{x}, \bar{\bbeta}^{\intercal} \boldsymbol{X}_t + \delta_1 \|\boldsymbol{X}_t\| \geq \kappa_1(\bbeta)  \right\} \supset \left\{  {\boldsymbol{X}}/{\bar{a}_t} \leq \boldsymbol{x} , \, \bbeta_t^\intercal\boldsymbol{X}_t \geq  \kappa_1(\bbeta)\right\}.
\end{align*}
Define the sequence of sets:
\[
G_n = \left\{ \boldsymbol{z} : \hat{\mathbf{a}}_{\mathbf{{1}/{n}}} \boldsymbol{z} \leq\boldsymbol{x} , \bar\bbeta^\intercal\boldsymbol{z} +{1}/{n} \|\boldsymbol{z}\| \geq \kappa_1(\bbeta) \right\}.
\]
Then, $G_n\downarrow G=\left\{ \boldsymbol{z}:\hat{\mathbf{a}}\boldsymbol{z} \leq\boldsymbol{x},\bar\bbeta^\intercal\boldsymbol{z}\geq \kappa_1(\bbeta)\right\}$. Now, using continuity of measure and the Portmanteau Theorem,
\begin{align*}
    \limsup_{t\to\infty}  t\Prob\left({\boldsymbol{X}}/{\bar{a}_t} \leq \boldsymbol{x} , \, \bbeta_t^\intercal\boldsymbol{X}_t \geq  \kappa_1(\bbeta)\right) \leq\mu(\hat{\mathbf{a}}_{\ddelta}\boldsymbol{z} &\leq \boldsymbol{x} \, , \bar\bbeta^\intercal\boldsymbol{z} + \delta_1\|\boldsymbol{z}\| \geq \kappa_1(\bbeta))\\
    &\leq \mu(\hat{\mathbf{a}}\boldsymbol{z} \leq \boldsymbol{x} \, , \bar\bbeta^\intercal\boldsymbol{z} \geq \kappa_1(\bbeta)) +\epsilon
\end{align*}
Now, $\limsup_{t\to\infty}  t\Prob\left( {\boldsymbol{X}}/{\bar{a}_t} \leq \boldsymbol{X} , \, \bbeta_t^\intercal\boldsymbol{X}_t \geq  \kappa_1(\bbeta)\right)$ is bounded above by    
$ \nu(\hat{\mathbf{a}}\boldsymbol{z} \leq \boldsymbol{X} \, , \bar\bbeta^\intercal\boldsymbol{z} \geq \kappa_1(\bbeta))$ (since $\epsilon>0$ was arbitrary). Hence, $\lim_{t\to\infty}t\Prob\left( {\boldsymbol{X}}/{\bar{a}_t} \leq \boldsymbol{x} , \, \bbeta_t^\intercal\boldsymbol{X}_t \geq  \kappa_1(\bbeta)\right)$ equals $\nu(\hat{\mathbf{a}}\boldsymbol{z} \leq \boldsymbol{x} \, , \bar\bbeta^\intercal\boldsymbol{z} \geq \kappa_1(\bbeta))$. Thus for all rectangles, $[\mathbf{0},\xx]$,  we have the convergence 
\[
\mathbb{P}\left( {\boldsymbol{X}}/{\bar{a}_t} \leq \xx \ \big \vert\  \bbeta_t^\intercal\boldsymbol{X}_t \geq  \kappa_1(\bbeta)\right) \to \nu\left(\mathbf{\hat{a}}\boldsymbol{z} \leq \xx, \bar\bbeta^\intercal\boldsymbol{z} \geq \kappa_1(\bbeta)\right).
\]
 Since  rectangles are a convergence determining class for $\Real^d$, convergence on rectangles implies convergence of measures (by Theorem 2.3, Example 2.3 from \cite{Billingsley}),  and the proof is complete.  \qed\\

\textbf{Proof of Lemma~\ref{lemma:UI}} Recall that a sequence of random variables, $X_t$ are uniformly integreble if there exists a $t_0$ such that $\lim_{K\to\infty}\sup_{t\geq t_0} E(X_t \mathbb{I}(X_t\geq K)) =0$.
A sufficient condition is that the existence of an $\epsilon>0$ such that $\sup_{t\geq t_0} E\vert X_t \vert^{1+\epsilon} <\infty$. To this end, for $\epsilon>0$, 
\begin{align}
    \Expc\left( \Bigg\|\frac{\boldsymbol{X}}{\bar{a}_t} \Bigg\|_2^{1+\epsilon} \, \Bigg\vert \, \bbeta^\intercal\boldsymbol{X}> \bar{a}_t\kappa_1(\bbeta)\right)&= \frac{\Expc\left( \Big\|\frac{\boldsymbol{X}}{\bar{a}_t} \Big\|_2^{1+\epsilon} \, \mathbf{I}\left( \bbeta^\intercal\boldsymbol{X}> \bar{a}_t\kappa_1(\bbeta)\right)\right)}{\Prob\left( \bbeta^\intercal\boldsymbol{X} \geq \Bar{a}_t \kappa_1(\bbeta)\right)}\label{eqn:first-b}\\
    &\leq \frac{\Expc\left( \left(\frac{\|\boldsymbol{X}\|_2}{\bar{a}_t}\right)^{1+\epsilon} \mathbf{I}( \|\bbeta\|_2\|\boldsymbol{X}\|_2> \bar{a}_t\kappa_1(\bbeta)\right)}{\Prob(\bbeta^\intercal\boldsymbol{X}> \bar{a}_t\kappa_1(\bbeta))}\label{eqn:second}.
\end{align}
Here, (\ref{eqn:first-b}) follows from the definition of conditional expectation. Lastly, notice that $\bbeta^{\intercal}\boldsymbol{X}\leq\|\bbeta\|_2\|\boldsymbol{X}\|_2$, and hence $\mathbf{I}\left(\bbeta^{\intercal}\boldsymbol{X}>\bar{a}_t\kappa_1(\bbeta)\right) \leq\mathbf{I}\left(\|\bbeta\|_2\|\boldsymbol{X}\|_2 >\bar{a}_t\kappa_1(\bbeta)\right)$ and (\ref{eqn:second}) follows.\\
It is easy to see that by the definition of $\kappa_1(\bbeta)$, the denominator of (\ref{eqn:second}) is $t^{-1}(1+o(1))$. We evaluate the numerator. For any non-negative function $g(\cdot)$, upon integration by parts, $\mathbf{E}(g(V)) = \int g^{'}(u)\bar{F}_V(u)du$. Set, $V=\|\boldsymbol{X}\|_2$, $g(u) = {u^{(1+\epsilon)}}$, and apply integration by parts to (\ref{eqn:second}):
\begin{align}\label{eqn:third}
  &{\Expc\left( \left(\frac{\|\boldsymbol{X}\|_2}{ \Bar{a}_t} \right)^{1+\epsilon} \mathbf{I}\left( \|\boldsymbol{X}\|_2> \frac{ \bar{a}_t\kappa_1(\bbeta)}{\|\bbeta\|_2}\right)\right)} \nonumber\\
  &= (1+\epsilon)\left(\frac{1}{\bar{a}_t\kappa_1(\bbeta)}\right)^{1+\epsilon}\int_{u\geq { \bar{a}_t\kappa_1(\bbeta)}/{{\|\bbeta\|_2}}} u^{\epsilon} \bar{F}(u)du.
\end{align}
Recall that $\xi=\max_{i=1}^{d}\xi_i$ and observe that $\bar{F}(\cdot)\in\mathrm{RV}_{-1/\xi}$, $U(x)=x^{\epsilon}\bar{F}(x)\in\mathrm{RV}_{\rho}$, where $\rho=\epsilon-1/\xi<-1$, for a sufficiently small $\epsilon$ (under the theorem hypothesis that $\xi<1$). Applying Karamata's Theorem (see Theorem B.1.5 in \cite{deHaan}), the right hand side of (\ref{eqn:third}) is bounded above for all sufficiently large $t$ by
\begin{equation}\label{eqn:Karamata-2}
    2(1+\epsilon)\|\bbeta\|_2^{(1+\epsilon)} \bar{F}\left( \bar{a}_t\kappa_1(\bbeta)/\|\bbeta\|_2\right)
\end{equation}
Plugging \eqref{eqn:Karamata-2} into \eqref{eqn:second}, for all $t$ sufficiently large,
\begin{align}\label{eqn:Karamata-3}
\Expc\left( \Bigg\|\frac{\boldsymbol{X}}{\bar{a}_t} \Bigg\|_2^{1+\epsilon} \, \Bigg\vert \, \bbeta^\intercal\boldsymbol{X}> \bar{a}_t\kappa_1(\bbeta)\right)\leq 2t \|\bbeta\|_2^{(1+\epsilon)} \bar{F}\left(\bar{a}_t\frac{\kappa_1(\bbeta)}{\|\bbeta\|}\right).
\end{align}
To conclude, 
\begin{lemma}\label{lemma:FinConv}
\begin{align}
\limsup_{t\to\infty}t\bar{F}\left(\bar{a}_t\frac{\kappa_1(\bbeta)}{\|\bbeta\|}\right) \leq \nu\left(\boldsymbol{z}:\|\hat{\mathbf{a}}\boldsymbol{z}\|_2 \geq\frac{\kappa_1(\bbeta)}{\|\bbeta\|}\right) <\infty\label{eqn:Karamata-4}
\end{align}
\end{lemma}

Then, from \eqref{eqn:Karamata-2}, \eqref{eqn:Karamata-3} and \eqref{eqn:Karamata-4}, for $t_0$ sufficiently large,
\[
\sup_{t\geq t_0} \Expc\left( \Bigg\|\frac{\boldsymbol{X}}{\bar{a}_t} \Bigg\|_2^{1+\epsilon} \, \Bigg\vert \, \bbeta^\intercal\boldsymbol{X}> \bar{a}_t\kappa_1(\bbeta)\right) \leq M,
\]
for an appropriately chosen, finite $M$, and we obtain the required uniform integrebility. \qed\\

\noindent \textbf{Proof of Lemma~\ref{lemma:FinConv}}

Integration with respect to the measure $\nu$ can be done conveniently in terms of the representation $\boldsymbol{Z} = \boldsymbol{R}/T^{\boldsymbol{\gamma}}$, where $\boldsymbol{R}$ is a $d$-dimensional random vector, and $T$ is an improper uniform random variable taking values in $[0,\infty)$ (see \cite{Rootzen} for a more detailed explanation for rewriting $\nu(\cdot)$ in terms of $\boldsymbol{R}$ and $T$). In particular, 
\begin{equation}\label{eqn:Kappa-beta-1}
    \iint_{\boldsymbol{r},t} \mathbb{I}\left({\frac{\bar{\bbeta}^\intercal\boldsymbol{r}}{t^{\boldsymbol{\gamma}}} >u}\right) dF(\boldsymbol{r})d{t}  
\end{equation}
Recall that $\bar{\bbeta} = \hat{\boldsymbol{a}} \bbeta$, and notice that $\hat{\mathbf{a}}_i>0$ only if $\xi_i=\xi$, where $\xi=\max_{i=1}^{d} \xi_i$ is the index of the fattest tail. Therefore for any fixed $u$, \eqref{eqn:Kappa-beta-1} can be re-written as
\begin{equation}\label{eqn:Kappa-beta}
    \iint_{\boldsymbol{r},t} \mathbb{I}\left({\frac{\bar{\bbeta}^\intercal\boldsymbol{r}}{t^{{\xi}}} >u}\right) dF(\boldsymbol{r})d{t}  
\end{equation}
Using Fubini's Theorem, \eqref{eqn:Kappa-beta} equals \[\frac{1}{u^{1/{\xi}}}\int_{\boldsymbol{r}} (\bar\bbeta^\intercal\boldsymbol{r})^{{1}/{\xi}} dF(\boldsymbol{r}).\]
Notice that above quantity is continuous and decreasing in $u$. Therefore 
\begin{align*}
    \kappa_s(\bbeta)&=\inf \{u:\nu(\bar\bbeta^\intercal\boldsymbol{z}\geq u) \leq s\}\\
    &= \{u:\nu(\bar\bbeta^\intercal\boldsymbol{z}\geq u) =s\}
\end{align*}
Setting $s=1$, $\kappa_1(\bbeta) = \|\bar\bbeta^{\intercal}\boldsymbol{r}\|_{\frac{1}{\xi}}$. Specifically, we have $\kappa_1(\bbeta) >0$ for all $\bbeta \in \Real^d_+$. Now, $\bar{F}(\|\bbeta\|^{-1}_2 \bar{a}_t\kappa_1(\bbeta)) = \Prob(\|\boldsymbol{X}\|_2 \geq \bar{a}_t \|\bbeta\|^{-1}_2\kappa_1(\bbeta))$. Write this as 
\begin{align}\label{eqn:CircProb}
\Prob\left(  \|\hat{\mathbf{a}}_t\boldsymbol{X}_t\|_2 \geq \frac{\kappa_1(\bbeta)}{\|\bbeta\|_2}\right) = \Prob\left(  \|(\hat{\mathbf{a}}+\ddelta_t)\boldsymbol{X}_t\|_2 \geq \frac{\kappa_1(\bbeta)}{\|\bbeta\|_2} \right).   
\end{align}
Since $\hat{\mathbf{a}}_t\to\hat{\mathbf{a}}$, for every $\ddelta$, for all large enough $t$, $-\ddelta\leq \hat{\mathbf{a}}_t - \hat{\mathbf{a}} \leq \ddelta$. Thus, for all $t$ large enough $\hat{\boldsymbol{a}}+\ddelta_t < \hat{\boldsymbol{a}}_t+\ddelta$. Hence,
\begin{align*}
\left\{\|\hat{\mathbf{a}}_t\boldsymbol{z}\|_2  \geq  \frac{\kappa_1(\bbeta)}{\|\bbeta\|_2}\right\}\subset \left\{\|(\mathbf{\hat{a}} + \ddelta)\boldsymbol{z}\|_2      
\geq  \frac{\kappa_1(\bbeta)}{\|\bbeta\|_2}\right\}.
\end{align*}
Thus, 
\begin{align}\label{eqn:Contains}
\Prob\left(  \|\hat{\mathbf{a}}_t\boldsymbol{X}_t\|_2 \geq \frac{\kappa_1(\bbeta)}{\|\bbeta\|_2} \right)  \leq \Prob\left(  \|(\mathbf{\hat{a}}+\ddelta)\boldsymbol{X}_t\|_2 \geq \frac{\kappa_1(\bbeta)}{\|\bbeta\|_2}\right).
\end{align}
From the Portmanteau Lemma, $\limsup_{t\to\infty} t\Prob\left(  \|(\mathbf{\hat{a}}+\ddelta)\boldsymbol{X}_t\|_2 \geq \frac{\kappa_1(\bbeta)}{\|\bbeta\|_2}\right)$ is bounded above by   $ \nu\left(\|({\mathbf{\hat{a}}}+\ddelta)\boldsymbol{z}\|_2  \geq  \frac{\kappa_1(\bbeta)}{\|\bbeta\|_2}\right)$. Define the sets 
\[
L_n = \left\{ \boldsymbol{z}: \left\| \left(\mathbf{\hat{a}} + {\mathbf{1}}/{\mathbf{n}}\right) \boldsymbol{z}  \right\|  \geq  \|\bbeta\|^{-1}_2\kappa_1(\bbeta) \right\}.
\]
Observe that $L_n \downarrow \{\boldsymbol{z}: \left\| \mathbf{\hat{a}}\boldsymbol{z}  \right\|  \geq  \|\bbeta\|^{-1}_2\kappa_1(\bbeta)\}$. Further, since $\mathbf{0} \not\in L_1$, $\nu(L_n)<\infty$ for all $n$. Thus, from the continuity of measure, for every $\epsilon>0$, for $n$ sufficiently large, 
\[
\nu(L_n) - \nu(\left\| \mathbf{\hat{a}}\boldsymbol{z}  \right\|  \geq  \|\bbeta\|^{-1}_2\kappa_1(\bbeta)) <\epsilon.
\]
Now, fix an $\epsilon$, and choose $\ddelta$ in \eqref{eqn:Contains} so small that $\ddelta < \frac{\mathbf{1}}{\mathbf{n}}$. Combining \eqref{eqn:CircProb} and \eqref{eqn:Contains}, we have
\begin{align*}
 &\limsup_{t\to\infty}t\bar{F}\left( \frac{\kappa_1(\bbeta)}{\|\bbeta\|_2}\right)\leq \nu\left(\boldsymbol{z}:\|\hat{\mathbf{a}}\boldsymbol{z}\|_2 \geq\frac{\kappa_1(\bbeta)}{\|\bbeta\|_2}\right) + \epsilon.
\end{align*}
Since $\epsilon$ above was arbitrary, the proof is complete. \qed\\
\textbf{Proof of Lemma~\ref{lemma:Zero-mass}}
Recall that $\boldsymbol{Z} = \boldsymbol{R}/T^{\mathbf{\xi}}$. Now, observe that for any $u>0$, using the continuity of measure, $ \nu(\bar\bbeta^\intercal\boldsymbol{z} =u) = \lim_{n\to\infty}\nu(\bar\bbeta^{\intercal}\boldsymbol{z} \in (u-\delta_n,u+\delta_n))$, where $\delta_n \to 0$. From the previous arguments, $\nu(\bar\bbeta^{\intercal}\boldsymbol{z} \in (u-\delta_n,u+\delta_n))$ equals 
\[
\left(1/(u-\delta_n)^{1/\xi} - 1/(u+\delta_n)^{1/\xi}\right)\iint_{\mathbf{r} } (\bbeta^\intercal\mathbf{r})^{1/\xi} dF(\mathbf{r}).
\] 
Since the above quantity goes to 0 as $n\to\infty$, the proof is complete.\qed\\
Proposition~\ref{prop:Gen-loss} extends Theorem~\ref{thm:convergence} to a general loss.
\begin{proposition}\label{prop:Gen-loss}
Let $\ell(\cdot)$ be a loss function satisfying the condition $\ell^{\prime}(u) = c_1u^{\rho}+o(u^{\rho})$. Further suppose that the covariates $\boldsymbol{X}$ satisfy Assumption~\ref{assumption:GEV}, such that $\rho<\xi^{-1}-1$. Then, the sensitivity of the CVaR of $l(\cdot)$ satisfies which satisfes $\ell^{'}(u)=u^{\rho}(1+o(1))$, with $\xi$ replaced by $\xi(\rho+1)$.
\end{proposition}
The proof of Proposition~\ref{prop:Gen-loss} follows upon observing 
$$\nabla C_{\beta_0}(\bbeta)=\bar{a}_t^{\rho+1}\Expc\left(  \frac{\boldsymbol{X}}{\bar{a}_t} (\frac{(\bar\bbeta^\intercal \boldsymbol{X})^{\rho}}{\bar{a}_t^\rho} )\ \vert \bbeta^\intercal \boldsymbol{X}  \geq v_t(\bbeta)  \right)$$
plus smaller terms. Recall that from Proposition~\ref{prop:Conditional-Lim}, 
$\mathcal{L}\left(\frac{\boldsymbol{X}}{\bar a_t} \, \bigg \vert \, \bbeta_t^\intercal\boldsymbol{X}_t \geq  \kappa_1(\bbeta) \right) \xrightarrow{\mathcal{L}} \mu$. Observe that the mapping $f(\mathbf{x}) = \mathbf{x} (\bbeta^\intercal{\mathbf{x}})^{\rho}$ is continuous in $\mathbf{x}$. Thus, applying the mapping theorem (see \cite{Billingsley}, Theorem 2.7), 
\[
\mathcal{L}\left(\frac{\boldsymbol{X}}{\bar a_t} \left(\frac{\bbeta^\intercal \boldsymbol{X}}{\bar{a}_t}\right)^{\rho}\, \bigg \vert \, \bbeta_t^\intercal\boldsymbol{X}_t \geq  \kappa_1(\bbeta) \right) \xrightarrow{\mathcal{L}} \mu_1(\cdot), 
\] 
where $\mu_1(d\mathbf{y}) =  \nu(\hat{\mathbf{a}}\mathbf{z}(\bar{\bbeta}^\intercal{\mathbf{z}})^{\rho} \in d\mathbf{y}, \bar\bbeta^\intercal\mathbf{z} \geq\kappa_1(\bbeta))$. Finally, the condition $\rho<\xi^{-1}-1$ ensures uniform integrebility, and thus convergence of the conditional expectation to a limit. This implies that $\nabla C_\beta(\bbeta) \sim \bar{a}_t^{\rho+1} E\mathbf{Y}_1$, where $\mathbf{Y}_1$ has the distribution  $\mu_1(\cdot)$. Now, notice that since $\bar{a}_t \in \mathrm{RV}_{\xi}$, $\bar{a}_t^{\rho+1} \in \mathrm{RV}(\xi(\rho+1))$. Then, we have
\begin{align*}
    \nabla C_\beta(\bbeta) &= \bar{a}_{t}^{\rho+1}E\boldsymbol{Y}_1(1+o(1))\\
    &= \bar{a}_{t_0}(t/t_0)^{\xi(\rho+1)}E\boldsymbol{Y}_1(1+o(1)) \text{ since $\bar{a}_t^{\rho+1} \in \mathrm{RV}(\xi(\rho+1)) $}\\
    &=(\beta_0/\beta)^{\xi(\rho+1)}\nabla C_{\beta_0}(\bbeta)(1+o(1))
\end{align*}\qed\\
\textbf{Proof of Corollary~\ref{thm:Consistency}}
 Recall that if $X_{n}$ be a sequence of random variables such that $X_{n}=a+O_{\mathcal{P}}(r_n)$ for some $r_n\to 0$. Then, for any once continuously differentiable function $f(\cdot):\Real\to\Real$, $f(X_{n})-f(a) = (X_n-a) f^{'}(a) + o_{\mathcal{P}}(r_n)$ . Now, with $X_n = (\hat{\xi}_n - \xi) \ln\frac{t_n}{t_{n,0}}$ and $f(x)=\mathrm{e}^x$, this gives,
\[
\left(\frac{t_n}{t_{n,0}}\right)^{\hat{\xi_n}} = \left(\frac{t_{n}}{t_{n,0}}\right)^\xi \left( 1+ (\hat{\xi}_n - \xi) \ln\frac{t_n}{t_{n,0}} \right)+o_{\mathcal{P}}(1).
\]
Since under the corollary assumptions, $l(\bbeta^\intercal\boldsymbol{X})$ is regularly varying,  $\hat{\xi}_n -\xi =o_{\mathcal{P}}(1)$ (see Section 2.3 of \cite{deHaan}),  
$$\left(\frac{t_n}{t_{n,0}}\right)^{\hat{\xi}_n} = \left(\frac{t_n}{t_{n,0}}\right)^{\xi}(1+o_{\mathcal{P}}(1)).$$
From the corollary hypothesis, $\hat \hat{C}_{\beta_{0}(n)}(\bbeta) = C_{\beta_{0}(n)}(\bbeta)(1+o_{\mathcal{P}}(1))$. Thus, we have, 
\begin{align}\label{eqn:4}
\tilde{C}_{\beta(n)}(\bbeta) &=   \hat{C}_{\beta_{0}(n)}(\bbeta) \left(\frac{\beta_{0}(n)}{\beta({n})}\right)^{\xi}(1+o_{\mathcal{P}}(1)).\nonumber\\
& = C_{\beta_{0}(n)}(\bbeta)\left(\frac{\beta_{0}(n)}{\beta({n})}\right)^{\xi}(1+o_{\mathcal{P}}(1))\nonumber\\
&= C_{\beta(n)}(\bbeta) (1+o_{\mathcal{P}}(1)).
\end{align}
where \eqref{eqn:4} follows from Theorem~\ref{thm:convergence}, and since $\frac{\beta_{0}(n)}{\beta(n)} < K_1$, and $\beta_n\to 0$. \qed\\


\textbf{Proof of \eqref{eqn:Lim_Expc}:}
Observe that $\Expc\left((\bbeta^\intercal\boldsymbol{X}  - u)^{+} \right)^2$ equals
\begin{equation}\label{eqn:5}
\int_{\bbeta^\intercal \boldsymbol{X}\geq u} (\bbeta^\intercal\boldsymbol{X}  - u) ^2 f(\xx)d\xx.
\end{equation}
Let $u_t=u^{\alpha_{*}}$, and define $ \boldsymbol{p} = u_t^{1/\alpha}\boldsymbol{X}$.  Therefore, \eqref{eqn:5} equals
\[  
u^2\prod_{i=1}^{d} u_{t}^{1/\alpha_i}  \cdot \int_{\bbeta_t^\intercal \boldsymbol{p} \geq 1}  (\bbeta_t^\intercal\boldsymbol{p}  - 1) ^2 f(u_t^{1/\alpha_i}\boldsymbol{p})d\boldsymbol{p}.
\]
As before, this is upper bounded by $1+\epsilon$ times
\[
u^{2-\alpha_*}L(u^{\alpha^{*}}) \int_{\bbeta_t^\intercal \boldsymbol{p} \geq 1}  (\bbeta_t^\intercal\boldsymbol{p}  - 1) ^2\Psi(\boldsymbol{p})d\boldsymbol{p} = u^{2-\alpha_*}L(u^{\alpha_*})\kappa(\bbeta)(1+o(1)).
\]
A matching lower bound may be established, which completes the proof\qed.

\begin{thebibliography}{9}
\bibitem{Ban}
Ban, G. Y., El Karoui, N., \& Lim, A. E. (2018). Machine learning and portfolio optimization. Management Science, 64(3), 1136-1154.
\bibitem{Billingsley}
  Billingsley, P. (2013). \textit{Convergence of probability measures.} John Wiley \& Sons.
\bibitem{Lam} Blanchet, J. and Lam, H., 2012. \textit{State-dependent
  importance sampling for rare-event simulation: An overview and
  recent advances}. Surveys in Operations Research and Management
  Science, 17(1), pp.38-59.
\bibitem{Bucklew}
  Bucklew, J. (2013).\textit{ Introduction to rare event simulation}. Springer Science \& Business Media.
\bibitem{Caccioli18}
  Caccioli, F.,  Kondor I., \&  Papp G. (2018)
  \textit{Portfolio optimization under Expected Shortfall: contour maps of
    estimation error}. Quantitative Finance, 18:8, 1295-1313.
\bibitem{Chow}
Chow, Y., Tamar, A., Mannor, S. and Pavone, M., 2015. \textit{Risk-sensitive and robust decision-making: a cvar optimization approach}. In Advances in Neural Information Processing Systems (pp. 1522-1530).
\bibitem{deHaan}
De Haan, L., \& Ferreira, A. (2007). \textit{Extreme value theory: an introduction}. Springer Science \& Business Media.
\bibitem{deHaanRV}
De Haan, L., \& Resnick, S. (1987). \textit{On regular variation of probability densities}. Stochastic processes and their applications, 25, 83-93.
\bibitem{Glasserman}
Glasserman, P., Heidelberger, P., \& Shahabuddin, P. (2002). Portfolio value‐at‐risk with heavy‐tailed risk factors.\textit{ Mathematical Finance}, 12(3), 239-269.
\bibitem{HongVaR}
Sun, L., \& Hong, L. J. (2010). Asymptotic representations for importance-sampling estimators of value-at-risk and conditional value-at-risk.\textit{ Operations Research Letters}, 38(4), 246-251.
\bibitem{HongCVaR}
Hong, L. J., \& Liu, G. (2009). Simulating sensitivities of conditional value at risk. \textit{Management Science}, 55(2), 281-293.
\bibitem{Krokhmal}
Krokhmal, P., Palmquist, J., \& Uryasev, S. (2002). Portfolio optimization with conditional value-at-risk objective and constraints. \textit{Journal of risk}, 4, 43-68.
\bibitem{Lim}
Lim, A. E., Shanthikumar, J. G., \& Vahn, G. Y. (2011). Conditional value-at-risk in portfolio optimization: Coherent but fragile. \textit{Operations Research Letters}, 39(3), 163-171.
\bibitem{Mainik}
Mainik, G., \& Rüschendorf, L. (2010). On optimal portfolio diversification with respect to extreme risks. \textit{Finance and Stochastics}, 14(4), 593-623
\bibitem{mcneil}
  McNeil, A. J., Frey, R., \& Embrechts, P. (2015). Quantitative risk management: concepts, techniques and tools-revised edition. Princeton university press.
\bibitem{Resnick}
Resnick, S. I. (2007). \textit{Heavy-tail phenomena: probabilistic and statistical modeling}. Springer Science \& Business Media.
\bibitem{Resnickbook}
Resnick, S. I. (2013). \textit{Extreme values, regular variation and point processes}. Springer.
\bibitem{Rockafellar}
Rockafellar, R. T., \& Uryasev, S. (2000). Optimization of conditional value-at-risk. \textit{Journal of risk}, 2, 21-42.
\bibitem{Rootzen}
Rootzén, H., Segers, J., \& Wadsworth, J. L. (2018). Multivariate peaks over thresholds models. \textit{Extremes}, 21(1), 115-145.
\bibitem{Scaillet}
Scaillet, O. (2004). Nonparametric estimation and sensitivity analysis of expected shortfall.\textit{ Mathematical Finance: An International Journal of Mathematics, Statistics and Financial Economics}, 14(1), 115-129.
\bibitem{Shapiro}
Ruszczyński, A., \& Shapiro, A. (2006). Optimization of convex risk functions. \textit{Mathematics of operations research, 31(3)}, 433-452.
\bibitem{Tamar}
Tamar, A., Glassner, Y.,\& Mannor, S. (2015, February). Optimizing the CVaR via sampling. \textit{In Twenty-Ninth AAAI Conference on Artificial Intelligence}.
\bibitem{Williamson}
Williamson, R. C., \& Menon, A. K. (2019).\textit{ Fairness risk measures}. arXiv preprint arXiv:1901.08665.
\end{thebibliography}
\end{document}